\documentclass[twocolumn]{aastex631}

\newcommand{\msolar}{M$_{\odot}$}
\newcommand{\lsolar}{L$_{\odot}$}
\usepackage{CJK}
\usepackage[whole]{bxcjkjatype} 
\usepackage{CJKutf8}
\usepackage{rotating}
\usepackage{longtable}
\usepackage{ulem}
\usepackage{soul}

\begin{document}

\title{The environmental dependence of mid-IR luminous dusty Supernovae}

\author[0000-0002-6986-5593]{Lin Xiao}
\altaffiliation{corresponding author: linxiao@hbu.edu.cn}
\affiliation{Department of Physics, Hebei University, Baoding, 071002, China}
\affiliation{Hebei Key Laboratory of High-precision Computation and Application of Quantum Field Theory, Baoding, 071002, China}
\affiliation{Hebei Research Center of the Basic Discipline for Computational Physics, Baoding, 071002, China}

\author[0009-0008-5182-5033]{Zeyue Peng}
\affiliation{Department of Physics, Hebei University, Baoding, 071002, China}
\affiliation{Hebei Key Laboratory of High-precision Computation and Application of Quantum Field Theory, Baoding, 071002, China}
\affiliation{Hebei Research Center of the Basic Discipline for Computational Physics, Baoding, 071002, China}

\author[0000-0002-1296-6887]{Llu\'is Galbany}
\affiliation{Institute of Space Sciences (ICE, CSIC), Campus UAB, Carrer de Can Magrans, s/n, E-08193 Barcelona, Spain}
\affiliation{Institut d’Estudis Espacials de Catalunya (IEEC), E-08034 Barcelona, Spain}

\author[0000-0003-4610-1117]{Tam\'as Szalai}
\affiliation{Department of Experimental Physics, Institute of Physics, University of Szeged, H-6720 Szeged, Dóm tér 9, Hungary}

\author[0000-0003-2238-1572]{Ori D. Fox}
\affiliation{Space Telescope Science Institute, 3700 San Martin Drive, Baltimore, MD 21218, USA}

\author[0000-0001-7201-1938]{Lei Hu}
\affiliation{McWilliams Center for Cosmology, Department of Physics, Carnegie Mellon University, 5000 Forbes Ave, Pittsburgh, 15213, PA, USA}

\author[0000-0003-3031-6105]{Maokai Hu}
\affiliation{Physics Department, Tsinghua University, Beijing 100084, China}

\author[0000-0001-6540-0767]{Thallis Pessi}
\affil{European Southern Observatory, Alonso de Córdova 3107, Vitacura, Casilla 19001, Santiago, Chile}

\author[0000-0002-6535-8500]{Yi Yang
\begin{CJK}{UTF8}{gbsn}
(杨轶)
\end{CJK}}
\affiliation{Physics Department and Tsinghua Center for Astrophysics (THCA), Tsinghua University, Beijing, 100084, China}
\affiliation{Department of Astronomy, University of California, Berkeley, CA 94720-3411, USA}

\author[0000-0003-1169-1954]{Takashi J. Moriya}
\affiliation{National Astronomical Observatory of Japan, National Institutes of Natural Sciences, 2-21-1 Osawa, Mitaka, Tokyo 181-8588, Japan}
\affiliation{Graduate Institute for Advanced Studies, SOKENDAI, 2-21-1 Osawa, Mitaka, Tokyo 181-8588, Japan}
\affiliation{School of Physics and Astronomy, Monash University, Clayton, Victoria 3800, Australia}

\author[0000-0001-9204-7778]{Zhanwen Han}
\affiliation{Yunnan Observatories, Chinese Academy of Sciences, Kunming 650216, China}
\affiliation{Key Laboratory for the Structure and Evolution of Celestial Objects, Yunnan Observatories, CAS, Kunming 650216, China}
\affiliation{International Centre of Supernovae, Yunnan Key Laboratory, Kunming, 650216, China}

\author[0000-0002-7334-2357]{Xiaofeng Wang}
\affiliation{Physics Department and Tsinghua Center for Astrophysics (THCA), Tsinghua University, Beijing, 100084, China}
\affiliation{Beijing Planetarium, Beijing Academy of Science and Technology, Beijing, 100044, China}

\author[0009-0004-4256-1209]{Shengyu Yan}
\affiliation{Physics Department and Tsinghua Center for Astrophysics (THCA), Tsinghua University, Beijing, 100084, China}



\begin{abstract}
Using the {\it Spitzer} and {\it WISE} images, we discovered 42 mid-IR luminous dusty supernovae with local integral-field spectroscopy data. The observed mid-IR emission indicates the presence of newly formed dust, or pre-existing dust heated by the radiation from the supernovae or circumstellar medium interactions. We carried out a systematic analysis of the supernova host environments and their dust properties, for understanding the dust-veiled exploding stars, and whether such an intense dust production process is associated with their local environments. We find that dusty supernovae prefer the locations with higher EW(H$\alpha$), lower metallicity, and heavier host extinctions compared to typical SN types, and they show the same increasing sequence in the values of EW(H$\alpha$) and oxygen abundance from hydrogen-rich, type IIn and hydrogen-poor dusty supernovae. These differences in environmental properties of different dusty SN types indicate the diversity of their progenitors. We also found that one marginal correlation is a negative correlation between the SN dust mass and star formation rate. This means that SNe would be more mid-IR luminous and more dust-rich at the region with lower star formation rate. However, the SN dust mass show no correlation with the metallicity and the host extinction, which were thought to be key factors affecting the mass-loss history of progenitors and the CSM environment of SNe. Therefore, the dust formation process in SNe might be insensitive to metallicity and the dust condition of their host environments.

\end{abstract}



\section{Introduction} \label{sec:intro}

The mid infrared (mid-IR) capability offered by the {\it Spitzer Space Telescope} ($Spitzer$ hereafter), systematic searches of dust in supernovae (SNe) have been carried out over the past decade (see, \citealp{Szalai2019,Szalai2021} for reviews). Spitzer Infrared Intensive Transients Survey (SPIRITS), provided a systematic mid-IR study of SNe of various types within nearby galaxies \citep{Tinyanont2016,Kasliwal2017}. Most of the dust emission in SNe can be attributed to being either pre-existing or newly formed dust heated by later-time circumstellar medium (CSM) interactions \citep{Fox2011, Szalai2013}. The inferred dust mass is critical to determine whether SN is playing a major role in enriching the dust budget within the interstellar medium (ISM). The dust grain species and the radial distribution profile would provide an immediate trace of the progenitor's pre-explosion mass-loss history, thus constraining the progenitor system \citep{Matsuura2009,Fox2011,Fox2013,Fransson2014,Tinyanont2019}. 

Characterizations of the dust present in such dusty SNe have revealed the presence of a substantial amount of dust over a broad range of SN subtypes. Type IIn SNe were the original class of SNe exhibiting clear signatures of interaction with the surrounding CSM. Progressively, signs of SN interactions were found in more ``normal" SNe and dust emission became dominated at late-time ($>$ 100days) photometry. \cite{Meikle2011,Szalai2011,Andrews2016,Tinyanont2016} found a mid-IR rebrightening due to warm dust emission in type IIP SNe, which were also theoretically thought to be the best candidates for dust formation among SNe due to its thicker outer envelope \citep{Kozasa2009,Gall2011}. Some stripped-envelope SNe (including type Ib/c, Ibn, and IIb) also show mid-IR emission at late times. The well-known case is SN 2014C which transformed from type Ib to an interacting type IIn over the course of a year \citep{Milisavljevic2015,Margutti2017}, and there is a  strong and long-lasting interaction observed over thousands of days post explosion by {\it Spitzer} \citep{Tinyanont2019}. Such a transformation from a H-poor SN Ib/c to a H-rich interacting SN IIn had also been observed in SN 2001em \citep{Chugai2006} and SN 2004dk \citep{Mauerhan2018}. Additionally, Type Ia including its subtypes exploding in dense, H-rich CSM shells, can also be mid-IR bright with significant dust component  \citep{Silverman2013,Fox2015,Graham2017}. A subluminous thermonuclear Type Iax SN 2014dt showed an excess of mid-IR emission at about one year post explosion \citep{Fox2016}, and a later ($> +310$ days) rebrightening in mid-IR was observed in the type Ia-CSM SN 2018evt \citep{Wang2024}. \cite{Siebert2024} found a strong thermal dust emission was present in the “super-Chandrasekhar” mass type Ia SN 2022pul, suggesting for a SN exploding within CSM.

The ability of strong CSM interaction and significant dust production enables dusty SNe to heavily influence their immediate environments and the evolution of their host galaxies. The study of the SN environments, although not a direct approach, has been able to provide independent constraints on the properties of SN progenitor systems \citep{Anderson2015,Chen2017,Galbany2018,Kuncarayakti2018,Xiao2019}. Many SN environment analyses have used the capabilities of integral-field spectroscopy (IFS) to investigate the SN associated HII regions within host galaxies for a more precise study of the SN host environment. Recently, \cite{Pessi2023} using data from the Very Large Telescope (VLT) with the Multi Unit Spectroscopic Explorer (MUSE) instrument studied 112 CCSNe, and they found that stripped-envelope SNe (Ib and Ic) are located in environments with a higher median SFR and oxygen abundance than Type II. Based on the SN environmental study, \cite{Pessi2023b} investigate the dependence of CCSNe production on metallicity, and conclude that there is a strong decrease in the number of CCSNe per unit of star-formation as oxygen abundance increases. \cite{Xi2025} also studied the effect of metallicity in evolution of massive stars and their final core-collapse supernova (CCSN) explosions, but they found that metallicity plays a minor role in the origin of different CCSN types. \citep{Moriya2023} studied the correlation of SN optical properties with their host environments in a sample of 21 Type IIn SNe, and they find that type IIn SNe with a higher peak luminosity are preferentially located in environments with lower metallicity and younger stellar populations. 


A systematic characterization of dusty SN environments in different SN types could be of critical importance to understand the properties of the dust-veiled exploding stars, and whether SN dust formation is associated with the local environment of the progenitor populations. This is the main motivation of this first environment study of dusty SNe in the current work. This paper is organized as follows: Section \ref{sec:Observations} describes the selection of our dusty SNe, their mid-IR photometry of {\it Spitzer} and {\it WISE}, and the dust SED fittings. Section \ref{sec:result} makes a statistical analysis of the environmental properties of dusty SNe. Section \ref{sec:dependence} investigates possible correlation between SN dust and their environmental proprieties. Finally, we conclude this paper in Section \ref{sec:conclusions}. \\\\\\\\

\section{Observations and data analysis} \label{sec:Observations}

\subsection{Collection of Supernova Data from the IFS sample}\label{sec:CCSNe-IFS}

We collected a sample of 275 nearby core-collapse supernovae (CCSNe) with IFS of their host galaxies from \cite{Galbany2018} (hereafter G18), \cite{Pessi2023} (hereafter P23) and \cite{Moriya2023} (hereafter M23). This sample consists of 154 hydrogen-rich SNe (including IIP, IIL, II), 83 hydrogen-poor SNe (including IIb, Ib, Ic), and 38 type IIn SNe. 

We note that our sample was taken from different works using different facilities. Firstly, G18 sample is constructed using 232 observed galaxies with IFS data from the PMAS/PPak Integral field Supernova hosts COmpilation (PISCO; \citealp{Galbany2018}), which is updated in terms of different SN subtypes including SN Ia and increased in the completeness of SN host galaxy sample with low-mass ($<10^{10}$\msolar) galaxies. So the host galaxies have $M_r$ varied from -24 to -12 mag and redshift up to 0.09 with median $\sim$ 0.017. Due to the spatial pixel (spaxel) size of PISCO at $1^{"}\times1^{"}$ (namely 1 arcsec$^{2}$ in area), a typical size of the HII region segregation in PISCO data can be a few hundreds of parsecs depending on distance \citep{Galbany2018}. This size is significantly larger than individual HII regions, and usually 1 to 6 HII regions were selected in data with that resolution \citep{Mast2014}. 

Then for P23 sample, they are selected from the SN catalogue of All-Sky Automated Survey for Supernovae (ASAS-SN) and then covered by the All-weather MUse Supernova Integral field Nearby Galaxies (AMUSING) survey \citep{Galbany2016a}. Their host galaxies have redshift up to 0.06 with median $\sim$ 0.011 and $M_B$ ranging from -23 to -16 mag. For AMUSING data, the spaxel size is $0.2^{"}\times0.2^{"}$, five times smaller than PISCO, which can zoom into SN parent stellar population better and decrease the contamination from nearby ones. M23 sample selected only type IIn host galaxies from the PISCO, AMUSING, and the Mapping Nearby Galaxies at APO (MaNGA; \citealp{Bundy2015}) surveys. The spaxel size of MaNGA is $0.5^{"}\times0.5^{"}$, in the middle of that of PISCO and AMUSING. As a whole, the HII region segmentation under the resolution of the three samples is larger than the physical scale of a real HII region, and the contamination from nearby stellar populations would affect our result using it as a SN parent stellar population. 

Our IFS sample is a collection of these three samples, which show a spread distribution in host galaxy magnitude and redshift, and provide an unbiased analysis of different types of SN environments. Table \ref{tab:Environmentdata} reports the local environmental properties of the dusty SNe we studied in this work. We measured the EW(H$\alpha$) which can be a good measurement of the stellar population age of SN host HII regions, the star formation rate intensity log$\Sigma_{\rm SFR}$ to trace the relation between star forming and SN explosion, the metallicity indicated by the oxygen abundance 12+log(O/H)$_{\rm D16}$ measured by \cite{Dopita2016}, and the E(B-V) of the SN host HII regions from the Balmer decrement given by \cite{Dominguez2013} to recognize any significant effect of extinction caused by ambient CSM or ISM, respectively. A more detailed description of the calculation of the SN local environmental properties can be found in \citet{Galbany2018}. 

\subsection{Mid-IR photometry from {\it Spitzer} and {\it WISE}} 
\begin{figure*}[ht!]
\centering
\includegraphics[width=0.97\columnwidth]{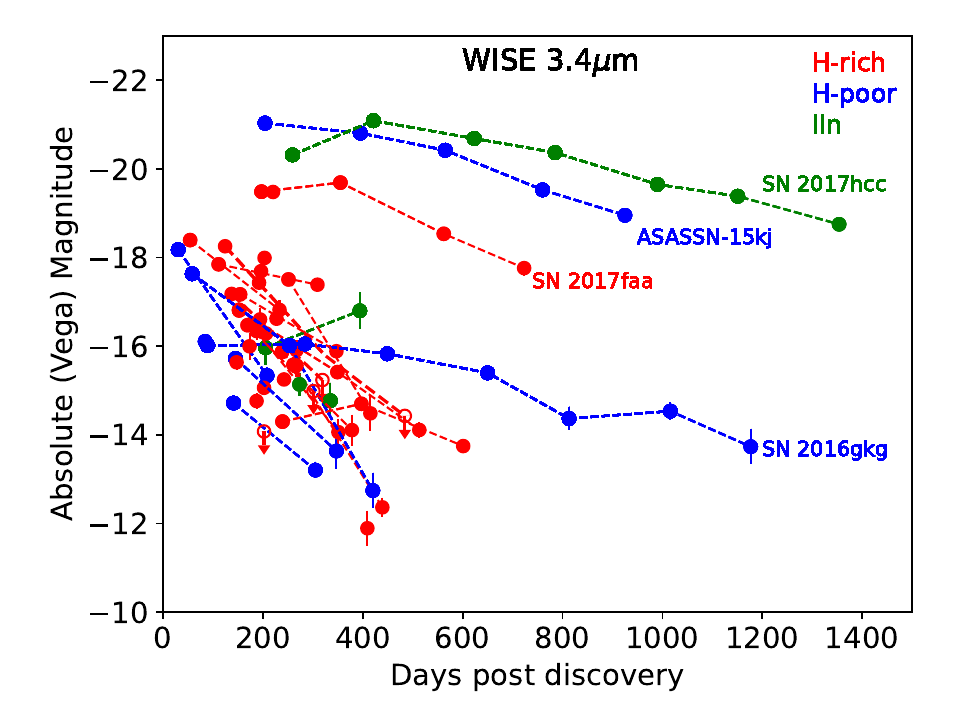}
\includegraphics[width=0.97\columnwidth]{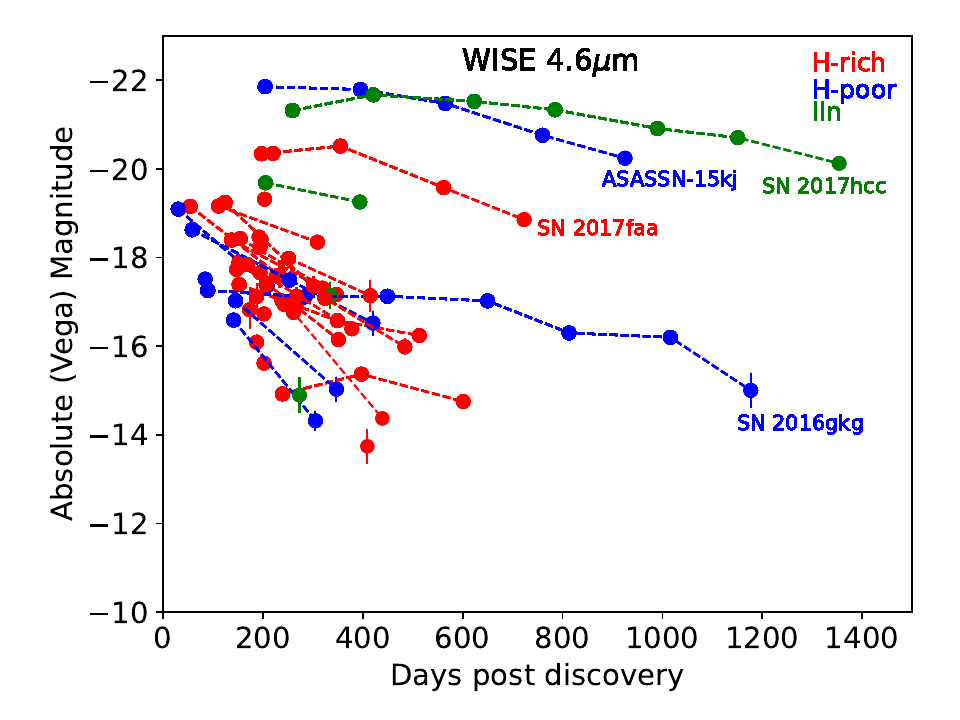}
\caption{The mid-IR photometry of all detected SNe at the band of 3.4 $\mu$m (left) and 4.6 $\mu$m (right), except that SN 2003gd and SN 2013ej detected by {\it Spitzer} are at 3.6 $\mu$m (right) and 4.5 $\mu$m (left). The colors are for different subclasses with hydrogen-rich SNe in red, hydrogen-poor SNe in blue and type IIn SNe in green. In the left panel, the empty symbols with downward arrows denote SNe whose absolute magnitudes were calculated with a 3$\sigma$ upper limits. All data are listed in Table \ref{tab:sndata-photo}.
\label{fig:mid-IR}}
\end{figure*}

The `dusty SNe' we studied are mid-IR luminous SNe especially at late time ($>$ 100 days). \citet{Szalai2019,Szalai2021} obtained the largest data set for over 100 SNe discovered to have late-time mid-IR emission. We cross-matched this {\it Spitzer} sample with the above PISCO/AMUSING/MaNGA SNe local environment sample and found only two SNe which are type IIP SNe SN 2003gd and SN 2013ej with identified dust measured.  

The Wide-field Infrared Survey Explorer ({\it WISE}; \citealp{Wright2010}) has mapped and kept going on monitoring the entire sky at mid-IR wavelengths of 3.4 $\mu$m (W1 band) and 4.6 $\mu$m (W2 band) every half year since its launch in 2010 except the operational gap between 2011 and 2013. Compared to {\it Spitzer}, {\it WISE} is inferior in spatial resolution with 6.1 and 6.6 arcsec in the W1 and W2 bands respectively, but superior in the sample size and monitoring duration \citep{Sun2022,Myers2024,Wang2024}. Therefore, we expect that more SNe in the above IFS sample could be observed by {\it WISE}. We performed a photometric analysis of the CCSNe in the above IFS sample using the time-resolved unWISE\footnote{\url{https://irsa.ipac.caltech.edu/data/WISE/unWISE/overview.html}} coadded images produced by \cite{Lang2014,Meisner2017a,Meisner2017b}. For each SNe, we created a reference image in each band by combining the pre-explosion images using SWarp \citep{Bertin2010}. For all the post-explosion images, we performed image subtraction with SFFT \citep{SFFT} to eliminate the flux contamination from the galaxy background. Then aperture photometry is performed on the subtracted images using Photutils \citep{larry_bradley_2023_7946442}. We used the standard aperture radius of 3 pixels and background annulus for the uncertainties from 18 to 25 pixels with 1 pixel of 2.75 arcsec following the explanatory supplement\footnote{\url{https://irsa.ipac.caltech.edu/data/WISE/docs/release/All-Sky/expsup/index.html}} of {\it WISE} data, and applied aperture corrections of 1.227 and 1.294 for 3.4 $\mu$m (W1 band) and 4.6 $\mu$m (W2 band), respectively. We set a 3$\sigma$ threshold for a positive detection in at least one {\it WISE} band. This 3$\sigma$ threshold in infrared SN studies was used by \cite{Fox2011} for {\it Spitzer} mid-IR photometry and has been widely adopted in works reviewed by \cite{Szalai2019}. For {\it WISE} photometry, the same 3$\sigma$ criterion was also employed by \cite{Sun2022} and \cite{Wang2024}. Finally, mid-IR emission was detected from 40 CCSNe, including 30 hydrogen-rich（H-rich) SNe, 7 hydrogen-poor (H-poor) SNe, and 3 type IIn SNe. We noted that there are four SNe (SN 2014ay, SN 2017gmr, SN 2018bbl, SN 2018cuf), which at some epochs are only positively detected at 4.6 $\mu$m, and we calculated a 3$\sigma$ upper limit at 3.4 $\mu$m. As discussed by \cite{Gutierrez2020}, SN2017ivv has a lower hydrogen mass than a typical SN II but more than a SN IIb, we included it in the group of H-poor SNe and labeled with II/IIb in Table \ref{tab:Environmentdata}.

Figure \ref{fig:mid-IR} plots the mid-IR photometry of all SNe with positive mid-IR detections including the two SNe detected by {\it Spitzer}. We note that the most SNe across all subtypes are detectable only within the first two years post discovery. These events are bright in mid-IR during the early-time CSM interaction, but the brightness declines quickly. We also highlight four specific targets: type II SN 2017faa, type Ic SNe ASASSN-15kj, type IIb SN 2016gkg, and type IIn SN 2017hpc, which produces a slowly declining mid-IR light curve in both 3.4 $\mu$m and 4.6 $\mu$m bands over two years and even up to 1000 days post discovery. Among the four SNe, Type IIn SN 2017hpc and type Ic SN ASASSN-15kj are extremely bright in mid-IR. For type IIn SNe, \cite{Fox2011,Fox2013} showed that the mid-IR radiation is more likely from pre-existing dust which is radiatively heated by optical emission generated by ongoing interaction between the SN forward shock and CSM. Type IIb SN 2016gkg, like SN 2001em and SN 2014C detected by {\it Spitzer} previously, seems to encounter a strong interaction with CSM at later times although this case is quite rare in H-poor SNe \citep{Chugai2006,Tinyanont2016}. SN 2017faa is the third luminous one which is relatively brighter than other hydrogen-rich SNe, while SN 2016gkg is much dimmer than other longer-lived SNe. The diversity in mid-IR evolution of CCSNe likely corresponds to the extent of pre-SN mass loss that constructed their surrounding CSM, but may also suggest different shock velocities, progenitors and explosion environment.

So far the highest redshift of CCSNe detected by {\it WISE} is a type II SN 2018hov of z $=$ 0.067 \citep{Myers2024}. Given the detection limit of {\it WISE} in terms of redshift $<$ 0.07, Figure \ref{fig:reshift-IFS} in appendix presents the redshift distribution of the IFS sample with the median redshift $\sim$ 0.015 (0.015 $\pm$ 0.008 in average), and only one type IIn SN 2017ghw of z $=$ 0.0762 in sample M23 lies beyond 0.07 which could be lost by the detection of {\it WISE}. In addition, we note that {\it WISE} observation scans the explosions since 2010 and have the most effective detection within the two years post SNe as shown in Figure \ref{fig:mid-IR}, so that around 50\% of CCSNe in the IFS sample occurred before 2008 could be dusty and mid-IR luminous but lost by {\it WISE}. Therefore, the SNe without positive detection by {\it Spitzer} and {\it WISE} could not be simply taken as non-dusty SNe. In this work, we would not discuss non-dusty SNe, and compared the mid-IR luminous dusty SNe detected by {\it Spitzer} and {\it WISE} with the whole IFS CCSN sample in terms of their environmental properties and in different typical SN types (H-rich, H-poor and type IIn) as discussed in Section \ref{sec:result}.\\

\subsection{SED Fitting and Dust properties}\label{SED-fitting}

\begin{figure*}[ht!]
\centering
\includegraphics[width=0.9\columnwidth]{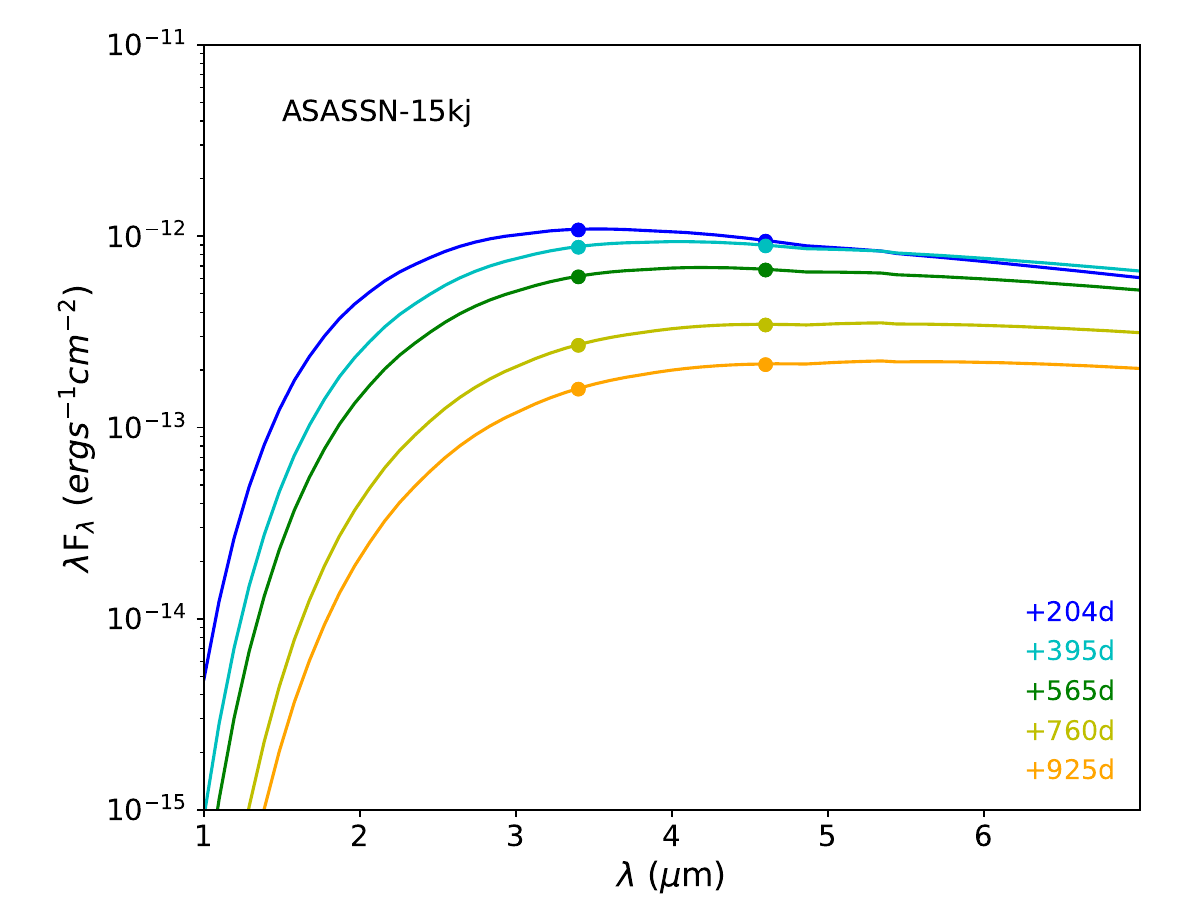}
\includegraphics[width=0.9\columnwidth]{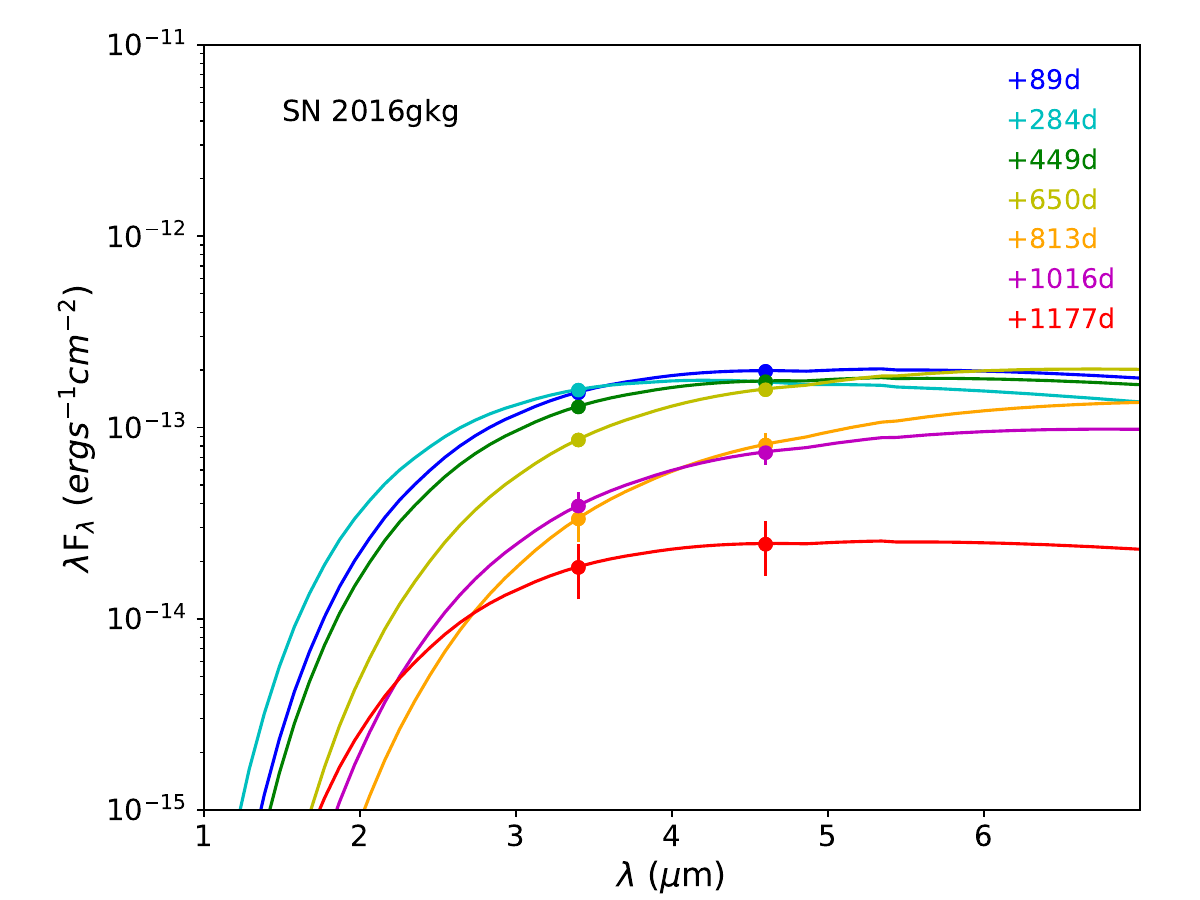}\\
\includegraphics[width=0.9\columnwidth]{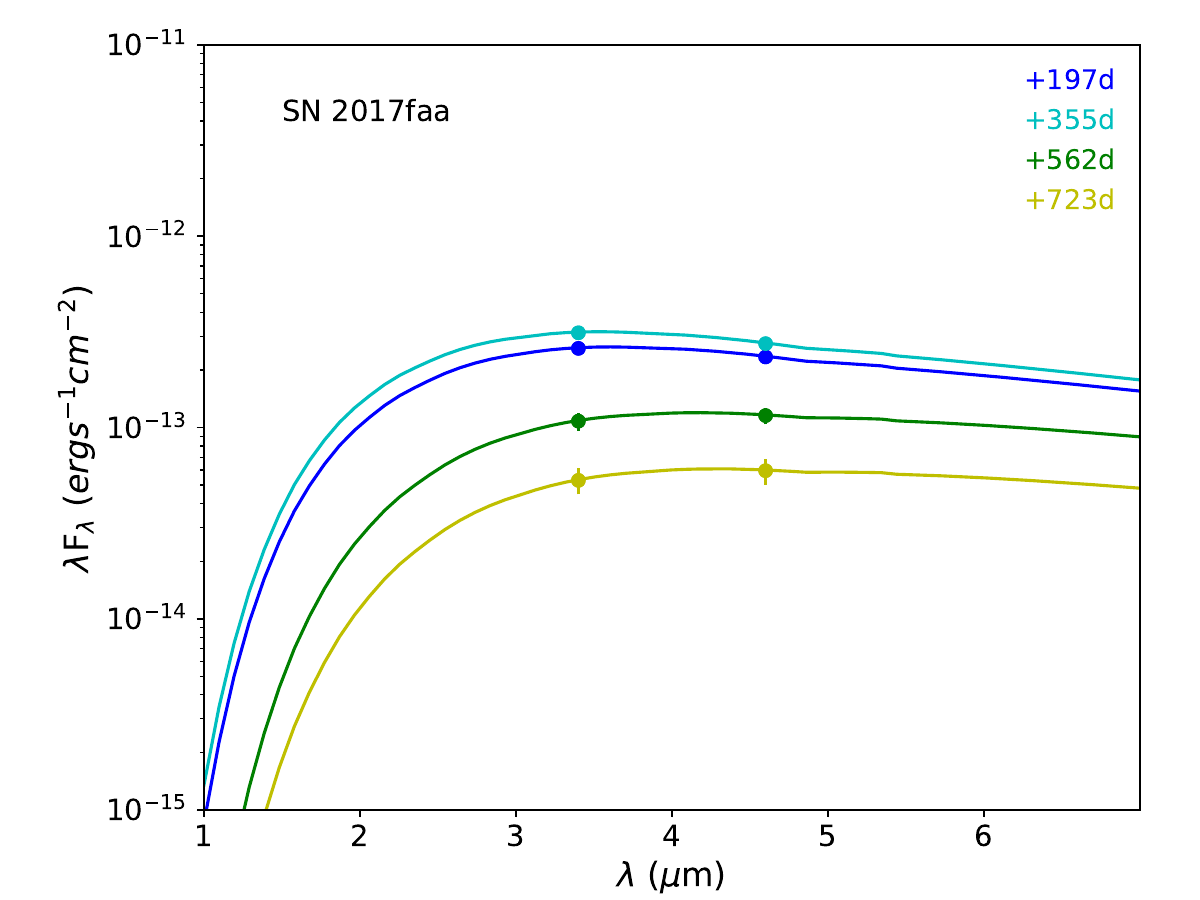}
\includegraphics[width=0.9\columnwidth]{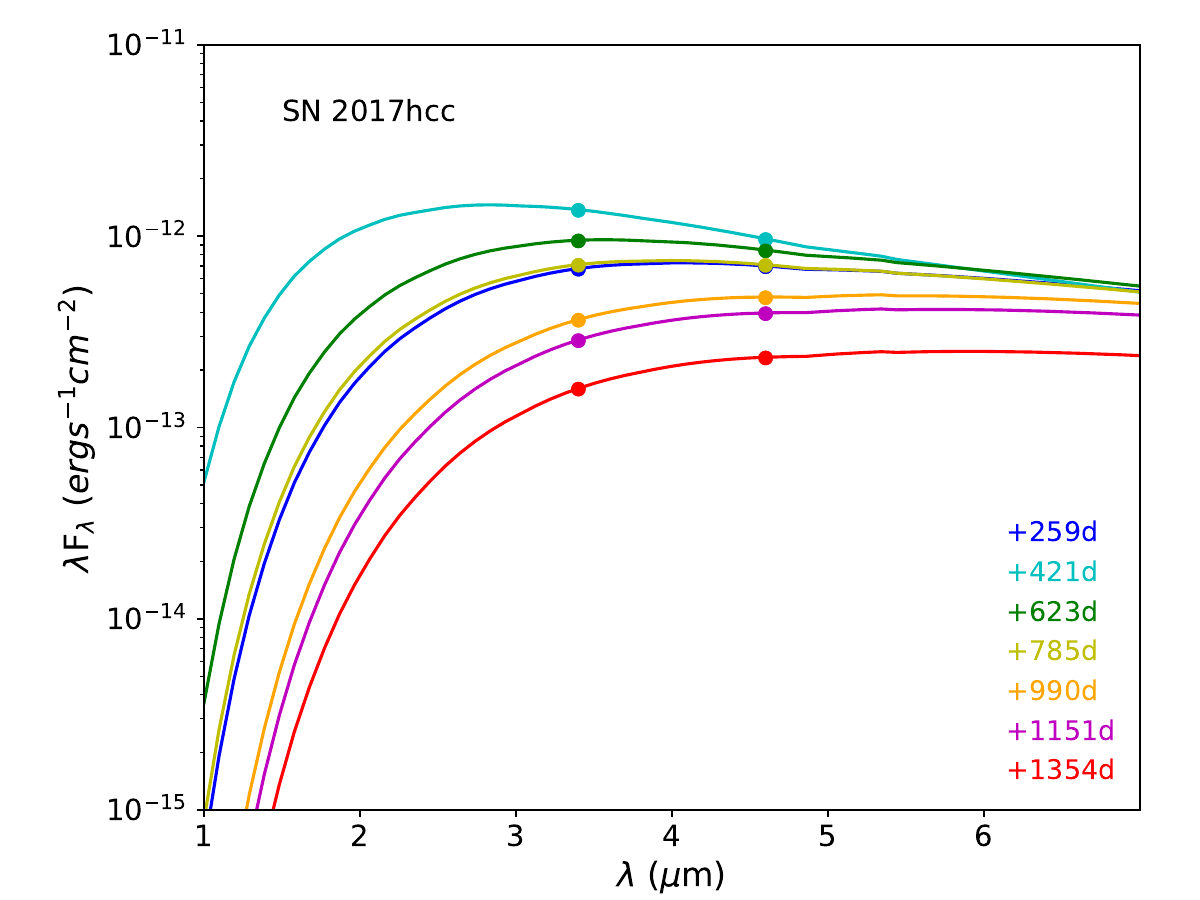}
\caption{SEDs of ASASSN-15kj, SN 2016gkg, SN 2017faa and SN 2017hpc at different epochs. {\it WISE} data are plotted in circles, and the overplotted lines are the best fit dust models following Equation（\ref{eq:flux}）. 
\label{fig:mid-SED}}
\end{figure*}

Mid-IR (3 $\leqslant$ $\lambda$ $\leqslant$ 25 $\mu$m) SEDs of SNe span the peak of thermal emission of dust with temperatures in the range of 100 $\lesssim$ $T_{\rm dust}$ $\lesssim$ 1000 K, providing useful constraints on the properties of dust \citep{Fox2011,Szalai2013}. We analyze the SNe following the method used in a number of papers  \citep{Fox2010,Fox2011,Szalai2013,Tinyanont2016,Shahbandeh2023} by assuming thermal emission only from an optically thin dust shell, whose spectrum is expressed as: 

\begin{equation}
    F_{\lambda}^{\rm obs} =\frac{ M_{dust}^{\rm obs} B_{\lambda}(T_{dust}) \kappa(a)}{D^2}\,
    \label{eq:flux}
\end{equation} 
\\
\noindent 
where $M_{dust}^{\rm obs}$ is the observed dust mass, $T_{dust}$ is the dust temperature, $B_{\lambda}(T_d)$ is the Planck function, $\kappa(a)$ is the dust mass absorption coefficient (see \citealp{Sarangi2022} for the values used in this work), $a$ is the grain radius, and $D$ is the distance to the source from the observer. Since {\it WISE} only has two photometry points at each epoch, we are limited to using one temperature component and one dust composition. We also choose a pure graphite composition with single-size grains of $a = 0.1 \mu$m following \cite{Fox2010,Fox2011}. We note that at early time ($\lesssim$ 200 days), the SN light still dominates the emission in mid-IR and the results from SED fitting in those epochs could only be an upper limit of the measured dust properties. We also note that the W1 (3.4 $\mu$m) band can be contaminated by the PAH emission and the W2 band (4.6 $\mu$m) band can be contaminated by carbon monoxide (CO) emission at 4.65 $\mu$m. 

We fit Equation (\ref{eq:flux}) to our data using the method \texttt{curve-fit} in \texttt{SciPy} package \citep{2020SciPy-NMeth}, which varies fitting parameters to find the best fit to the data using the method of least squares \citep{Jones2001}. Fig. \ref{fig:mid-SED} plots the best-fit dust models for the four longer-lived SNe (ASASSN-15kj, SN 2016gkg, SN 2017faa and SN 2017hpc). Table \ref{tab:sndata-photo} lists the dust parameters of all SNe from the best SED fitting. 

We also worked out the minimum shell size of an observed dust component, by assuming that the observed flux is a perfect blackbody (BB) as in \cite{Fox2011}, so that 

\begin{equation}
\label{eq:Rbb}
R_{\mathrm{BB}}(t) =\sqrt{ \frac{L}{4 \pi \sigma T^{4}} } = \sqrt{\frac{4 \pi d^{2} \int F^{\rm obs}_{\rm \lambda} d \lambda}{4 \pi \sigma T^{4}}}
\,
\end{equation}
\\
\noindent
where $\sigma$ is Stefan-Boltzmann's constant. The measured black body radius R$_{BB}$ at each epoch are also listed in Table \ref{tab:sndata-photo}. 

In addition, the shock radius $r_{s}$ $\sim$ $v_{s}t$, for a constant shock velocity $v_s$, defines the maximum radius that the forward shock can travel in a time t. The associated dust mass processed by the forward shock of the SN is given by Equation (6) in \cite{Fox2011}, assuming a dust to gas ratio of 0.01: 

\begin{equation}
\label{eq:shock}
M_{\mathrm{d}}({\rm M}_{\odot}) \approx 0.0028 \left( \frac{v_{\mathrm{s}}}{15,000 \,\mathrm{km} \,\mathrm{s}^{-1}} \right)^3 \left( \frac{t}{\mathrm{yr}} \right)^2 \left( \frac{a}{\mathrm{\mu m}}\right)\, 
\end{equation}
\\
\noindent where $v_\mathrm{s}$ is the shock velocity, $t$ is the time post explosion, and $a$ is the dust grain size. The shock velocity corresponds to the broad component of optical spectral line width, which is typically $\sim$ 5000 to 15000 km$s^{\rm -1}$ for CCSNe.

\begin{figure*}[ht!]
\centering
\includegraphics[width=\columnwidth]{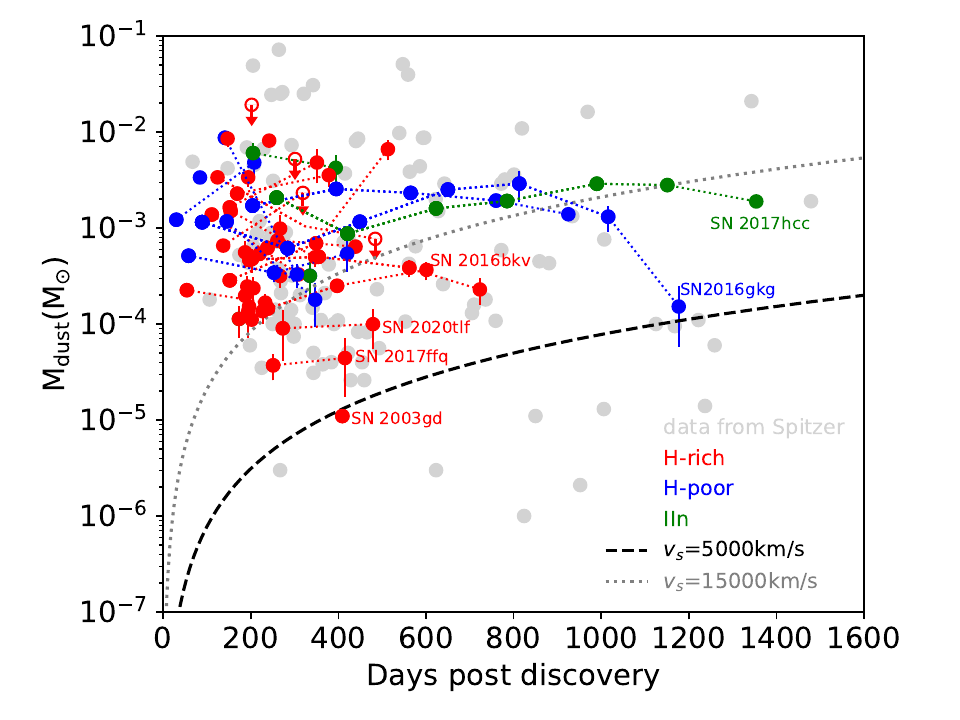}
\includegraphics[width=\columnwidth]{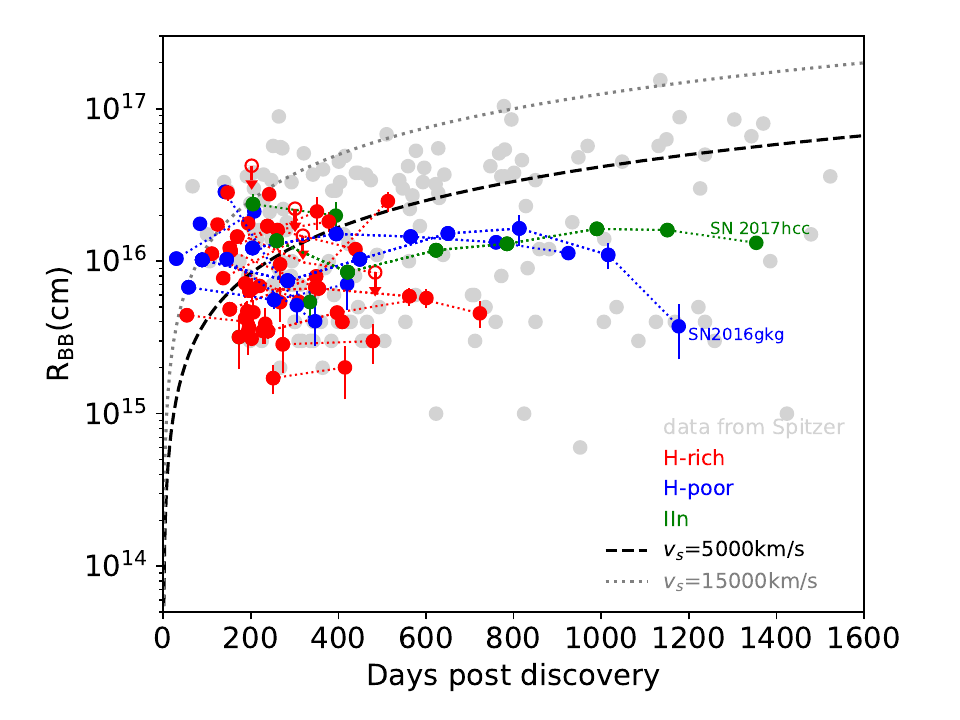}
\caption{Dust mass (M$_{\rm dust}$, left) and blackbody radius (R$_{\rm BB}$, right) of dusty SNe derived from the SED fittings. Filled circles denote the dusty SNe with H-rich SNe in red, H-poor SNe in blue and type IIn SNe in green, compared to the data from {\it Spitzer} \citep{Szalai2019} in the filled circles of light gray. The empty symbols with downward arrows denote a upper limit in the measured dust mass. Dotted lines on the left panel denote theoretical dust masses at shock velocities $v_{s} =$ 5000 and 15000km/s calculated by Equation (\ref{eq:shock}).
\label{fig:Dust-paras}}
\end{figure*}

The mid-IR emission has been explained with the presence of newly formed dust, or pre-existing dust heated by the radiation from the SN or CSM interactions \citep{Fransson2014,Fox2015}. As discussed in \cite{Tinyanont2016} and \cite{Szalai2019}, the shocked dust mass can distinguish these two scenarios as shown in Fig. \ref{fig:Dust-paras}. If the mass derived from SED fitting is larger than the predicted shocked dust mass even at the highest shock velocity of 15000 km$s^{\rm -1}$, the dust is pre-existing at the time of the explosion and radiatively heated by ongoing CSM interactions. In contrast, the dust is newly formed either in ejecta or a cool dense shell (CDS) behind the forward shock when R$_{\rm BB}$ is smaller than the shock radius $r_{s}$ at the lowest shock velocity of 5000 km$s^{\rm -1}$. While in cases where R$_{\rm BB}$ is between the shock radius $r_{s}$ at the shock velocity of 5000 km$s^{\rm -1}$ and 15000 km$s^{\rm -1}$, the dust radius is consistent with the shock front, suggesting that new dust may be forming in the CDS behind the forward shock but the presence of pre-existing dust should be still invoked to explain the amount of observed mid-IR radiations. Therefore, the origin of most SN dust in our sample are pre-existing dust dominated, and newly formed dust was definitely formed in 2003gd and SN 2016gkg at the last epoch considering the errors. 

We also noted that the SN dust varied from 10$^{-5}$ to 10$^{-2}$ ${\rm M}_{\odot}$ and can dominate the mid-IR emission of CCSNe from quite early time around 100 days up to 1000 days, with most of them clustered within 400 days post discovery which is consistent with the result of {\it Spitzer} \citep{Szalai2019}. The dust radius marked by R$_{\rm BB}$ extended from 10$^{15}$ to the order of 10$^{16}$ cm, sharing similar distribution with the result of {\it Spitzer}. However, the {\it Spitzer} sample presents a wider range of SN dust in both epochs and the amount of dust mass and radius. In addition, studies of \cite{Shahbandeh2023, Zsiros2024,Szalai2025} have also found a significant amount of newly formed dust in the ejecta even tens of years after explosion based on {\it JWST} data at longer wavelength. Therefore, the bias of the measured SN dust in our sample can be primarily due to wavelength limitation of {\it WISE} which is insensitivity to cold dust with $T_{\rm dust} \lesssim 400$K. Since we study the environmental properties of the dusty SNe by comparing them with the entire IFS sample, this bias has a limited impact on our results discussed in the following sections. On the other hand, the comparison of the SN dust within the H-rich, H-poor and type IIn dusty SNe suggests that these three SN types show no significant difference in SN dust mass, radius and epochs.

\begin{deluxetable*}{cccccccc}
\tablecaption{The local environmental properties of the 42 dusty SNe studied in this work. 
\label{tab:Environmentdata}}
\tablecolumns{5}
\tablehead{
\colhead{Name} & \colhead{Type} &  \colhead{Distance} &\colhead{EW(H$\alpha$)} & \colhead{12+log(O/H)$_{\rm D16}$} & \colhead{log$\Sigma_{\rm SFR}$} & \colhead{E(B-V)} & {References} \\
\colhead{ } & \colhead{ }  & \colhead{Mpc}  & \colhead{\AA} & \colhead{ } & \colhead{\msolar yr$^{-1}$kpc$^{-2}$} & \colhead{mag} 
}
\startdata
 \tableline
SN 2003gd	&	IIP &  8.97	&	43.18 	$\pm$	3.01 	&	8.49 	$\pm$	0.01 	&	-2.241 	$\pm$	0.021 	&	0.003 	$\pm$	0.121 	&	 \cite{VanDyk2003}\\
SN 2010dr	&	IIL	&	78.77	&	30.02 	$\pm$	2.96 	&	8.31 	$\pm$	0.02 	&	-2.282 	$\pm$	0.034 	&	0.255 	$\pm$	0.217 	&	 \cite{Kinugasa2010}\\
SN 2013ej	&	IIP	&	8.97	&	17.59 	$\pm$	1.20 	&	8.73 	$\pm$	0.02 	&	1.065 	$\pm$	0.021 	&	1.008 	$\pm$	0.727 	&	 \cite{Valenti2014}\\
SN 2014az	&	IIP 	&	59.89	&	44.77 	$\pm$	1.38 	&	8.01 	$\pm$	0.01 	&	-2.300 	$\pm$	0.033 	&	0.114 	$\pm$	0.227 	&	 \cite{Parker2014}\\
SN 2014ay	&	II	&	46.47	&	39.25 	$\pm$	1.17 	&	8.31 	$\pm$	0.02 	&	-1.873 	$\pm$	0.000 	&	0.358 	$\pm$	0.083 	&	 \cite{Holoien2014}\\
ASASSN-14az	&	IIb	&	31.34	&	25.48 	$\pm$	1.08 	&	7.86 	$\pm$	0.09 	&	-2.379 	$\pm$	0.000 	&	-			&	 \cite{Benetti2014} \\
SN 2014cw	&	II	&	40.23	&	-			&	8.09 	$\pm$	0.14 	&	-2.822 	$\pm$	0.000 	&	0.279 	$\pm$	0.288 	&	 \cite{Elias-Rosa2014} \\
ASASSN-14ma	&	IIP	&	58.54	&	64.85 	$\pm$	2.31 	&	8.10 	$\pm$	0.02 	&	-1.637 	$\pm$	0.000 	&	0.152 	$\pm$	0.041 	&	 \cite{Cao2014} \\
ASASSN-15ab	&	IIn	&	78.32	&	313.14 	$\pm$	1.23 	&	8.38 	$\pm$	0.01 	&	-3.347 	$\pm$	0.001 	&	0.130 	$\pm$	0.010 	&	 \cite{Shappee2015a} \\
ASASSN-15jp	&	II	&	42.01	&	48.08 	$\pm$	3.20 	&	8.28 	$\pm$	0.04 	&	-2.455 	$\pm$	0.000 	&	0.422 			&	 \cite{Falco2015} \\
ASASSN-15kj	&	Ic	&	82.38	&	44.94 	$\pm$	1.14 	&	8.61 	$\pm$	0.01 	&	-1.943 	$\pm$	0.000 	&	0.266 	$\pm$	0.045 	&	 \cite{Shappee2015b} \\
ASASSN-15ln	&	II	&	67.06	&	-			&	8.12 	$\pm$	0.10 	&	-2.342 	$\pm$	0.000 	&	0.202 	$\pm$	0.152 	&	 \cite{Benetti2015} \\
ASASSN-15ng	&	IIP	&	44.23	&	130.49 	$\pm$	3.25 	&	8.25 	$\pm$	0.01 	&	-1.269 	$\pm$	0.001 	&	0.235 	$\pm$	0.025 	&	 \cite{Shappee2015c}\\
SN 2015bj	&	II	&	50.04	&	40.11 	$\pm$	1.91 	&	8.35 	$\pm$	0.02 	&	-2.082 	$\pm$	0.000 	&	0.203 	$\pm$	0.070 	&	 \cite{Galbany2016b}\\
SN 2016B	&	IIP	&	19.38	&	86.01 	$\pm$	2.30 	&	7.90 	$\pm$	0.02 	&	-1.467 	$\pm$	0.001 	&	0.082 	$\pm$	0.024 	&	 \cite{Piascik2016}\\
SN 2016C	&	IIP	&	20.26	&	14.69 	$\pm$	1.16 	&	8.43 	$\pm$	0.04 	&	-2.738 	$\pm$	0.000 	&	-			&	 \cite{Sahu2016}\\
SN 2016X	&	IIP	&	18.05	&	58.98 	$\pm$	3.41 	&	8.01 	$\pm$	0.04 	&	-2.437 	$\pm$	0.000 	&	0.203 	$\pm$	0.100 	&	 \cite{Zheng2016}\\
SN 2016aqf	&	IIP	&	17.61	&	121.12 	$\pm$	4.16 	&	7.84 	$\pm$	0.03 	&	-1.139 	$\pm$	0.001 	&	0.098 	$\pm$	0.033 	&	 \cite{Ryder2016}\\
SN 2016bkv	&	IIP	&	8.79	&	6.54 	$\pm$	0.52 	&	8.77 	$\pm$	0.02 	&	-2.336 	$\pm$	0.024 	&	0.429 	$\pm$	0.204 	&	 \cite{Hosseinzadeh2018}\\
SN 2016blz	&	II	&	51.83	&	69.73 	$\pm$	2.00 	&	7.89 	$\pm$	0.04 	&	-1.424 	$\pm$	0.001 	&	0.294 	$\pm$	0.056 	&	 \cite{Magee2016a}\\
SN 2016bmi	&	II	&	33.12	&	15.88 	$\pm$	1.93 	&	8.19 	$\pm$	0.14 	&	-2.838 	$\pm$	0.000 	&	-			&	 \cite{Magee2016b}\\
SN 2016cdd	&	Ib	&	38.05	&	994.94 	$\pm$	20.23 	&	8.09 	$\pm$	0.03 	&	0.314 	$\pm$	0.152 	&	0.248 	$\pm$	0.036 	&	 \cite{Hosseinzadeh2016}\\
SN 2016cvk	&	IIn	&	47.81	&	24.37 	$\pm$	0.26 	&	8.60 	$\pm$	0.04 	&	-4.041 	$\pm$	0.010 	&	0.150 	$\pm$	0.020 	&	 \cite{Bersier2016a}\\
SN 2016cyx	&	II 	&	60.78	&	196.10 	$\pm$	5.39 	&	8.55 	$\pm$	0.01 	&	-1.025 	$\pm$	0.002 	&	0.165 	$\pm$	0.028 	&	 \cite{Falco2016}\\
SN 2016gkg	&	IIb	&	21.59	&	-			&	8.78 	$\pm$	0.06 	&	-3.600 	$\pm$	0.000 	&	-	$\pm$	0.384 	&	 \cite{VanDyk2016}\\
SN 2016hgm	&	II	&	33.56	&	148.06 	$\pm$	5.33 	&	8.36 	$\pm$	0.01 	&	-1.372 	$\pm$	0.001 	&	0.256 	$\pm$	0.033 	&	 \cite{Dimitriadis2016}\\
SN 2016hwn	&	II	&	96.42	&	19.86 	$\pm$	5.77 	&	8.28 	$\pm$	0.09 	&	-2.867 	$\pm$	0.000 	&	-			&	 \cite{Bersier2016b}\\
SN 2016iae	&	Ic	&	15.41	&	317.28 	$\pm$	13.39 	&	8.77 	$\pm$	0.03 	&	-0.895 	$\pm$	0.003 	&	1.019 	$\pm$	0.229 	&	 \cite{Jha2016} \\
SN 2017bzb	&	II	&	12.75	&	121.74 	$\pm$	18.71 	&	8.04 	$\pm$	0.05 	&	-2.425 	$\pm$	0.000 	&	0.066 	$\pm$	0.097 	&	 \cite{Morrell2017}\\
SN 2017faa	&	II	&	82.38	&	-			&	8.48 	$\pm$	0.01 	&	-1.931 	$\pm$	0.000 	&	0.229 	$\pm$	0.024 	&	 \cite{Somero2017} \\
SN 2017fbu	&	II	&	48.25	&	120.19 	$\pm$	6.59 	&	8.20 	$\pm$	0.02 	&	-2.173 	$\pm$	0.000 	&	0.067 	$\pm$	0.032 	&	 \cite{Rodriguez2017a} \\
SN 2017ffq	&	II	&	55.85	&	74.11 	$\pm$	1.91 	&	8.51 	$\pm$	0.01 	&	-0.948 	$\pm$	0.003 	&	0.315 	$\pm$	0.047 	&	 \cite{Rodriguez2017b} \\
SN 2017fqk	&	II	&	45.13	&	79.26 	$\pm$	3.84 	&	8.63 	$\pm$	0.02 	&	-2.132 	$\pm$	0.000 	&	0.419 	$\pm$	0.115 	&	 \cite{Tartaglia2017}\\
SN 2017gax	&	Ibc	&	20.71	&	84.55 	$\pm$	2.48 	&	8.85 	$\pm$	0.01 	&	-1.840 	$\pm$	0.000 	&	0.363 	$\pm$	0.036 	&	 \cite{Jha2017}\\
SN 2017gmr	&	IIP	&	22.03	&	57.65 	$\pm$	2.35 	&	8.42 	$\pm$	0.02 	&	-1.922 	$\pm$	0.000 	&	0.322 	$\pm$	0.055 	&	 \cite{Andrews2019}\\
SN 2017hcc	&	IIn	&	74.71	&	13.44 	$\pm$	0.59		&	7.88 	$\pm$	0.10 	&	-3.338 	$\pm$	0.000 	&	0.249 	$\pm$	0.232 	&	 \cite{Dong2017}\\
SN 2017hpi	&	II	&	24.69	&	-			&	8.39 	$\pm$	0.04 	&	-2.341 	$\pm$	0.000 	&	0.212 	$\pm$	0.139 	&	 \cite{Oneill2017}\\
SN 2017ivv	&	II/IIb$^{*}$	&	24.25	&	62.32 	$\pm$	6.13 	&	8.30 	$\pm$	0.10 	&	-2.659 	$\pm$	0.000 	&	0.058 	$\pm$	0.082 	&	 \cite{Gutierrez2020}\\
SN 2018pq	&	IIP	&	30.45	&	-			&	8.71 	$\pm$	0.01 	&	0.180 	$\pm$	0.213 	&	0.394 	$\pm$	0.044 	&	 \cite{Dubey2025}\\
SN 2018bbl	&	II	&	26.46	&	14.17 	$\pm$	1.08 	&	8.49 	$\pm$	0.06 	&	-2.975 	$\pm$	0.000 	&	0.236 	$\pm$	0.186 	&	 \cite{Cartier2018} \\
SN 2018cuf	&	II	&	46.47	&	48.97 	$\pm$	1.36 	&	8.67 	$\pm$	0.03 	&	-2.350 	$\pm$	0.000 	&	0.254 	$\pm$	0.074 	&	 \cite{Jha2018}\\
SN 2020tlf	&	IIP/L	&	37.56	&	115.73 	$\pm$	0.33 	&	8.23 	$\pm$	0.05 	&	-1.209 	$\pm$	0.031 	&	0.064 	$\pm$	0.033 	&	 \cite{Jacobson-Galan2022}\\
\tableline
\enddata
\tablecomments{All the supernova types, distances and environmental parameters are taken from the IFU sample of G18 \citep{Galbany2018}, P23 \citep{Pessi2023}  and M23 \citep{Moriya2023}. *As discussed by \cite{Gutierrez2020}, SN2017ivv has a lower hydrogen mass than a typical SN II but more than a SN IIb, we included it in the group of H-poor SNe and labeled with II/IIb. (This table is available in machine-readable form.)}
\end{deluxetable*}

\begin{deluxetable*}{ccccc}
\tablecaption{The median and mean value of the environmental properties of the three typical CCSN subtypes including all IFU samples and the dusty SN sample listed in Table \ref{tab:Environmentdata}. 
\label{tab:median4types}}
\tablecolumns{3}
\tablehead{
\colhead{Type} & \colhead{logEW(H$\alpha$)} & \colhead{12+log(O/H)$_{\rm D16}$} & \colhead{log$\Sigma_{\rm SFR}$} & \colhead{E(B-V)} \\
\colhead{ } & \colhead{\rm \AA} & \colhead{ } & \colhead{\msolar yr$^{-1}$kpc$^{-2}$} & \colhead{mag} 
}
\startdata
 \tableline
& & Median & & \\
\cline{2-5}
H-rich & 1.57 & 8.58 & -1.96 & 0.13 \\
H-poor & 1.56 & 8.59 & -1.89 & 0.15 \\ 
IIn & 1.58 &  8.57 & -2.47 & 0.20 \\
\tableline
\textbf{Dusty SNe} & \textbf{1.77} & \textbf{8.33} & \textbf{-2.21} & \textbf{0.24} \\
\textbf{H-rich} & 1.76 & 8.31 & -2.15 & 0.23 \\
\textbf{H-poor} & 1.89 & 8.61 & -1.94 & 0.27 \\ 
\textbf{IIn} & 1.82 &  8.38 & -3.55 & 0.16 \\
\tableline
& & Mean & & \\
\cline{2-5}
H-rich & 1.74$\pm$0.57 & 8.54$\pm$0.32 & -1.84$\pm$0.91 & 0.18$\pm$0.19 \\
H-poor & 1.66$\pm$0.58 & 8.58$\pm$0.26 & -1.86$\pm$0.66 & 0.17$\pm$0.16 \\ 
IIn & 1.76$\pm$0.66 &  8.50$\pm$0.29 & -2.73$\pm$1.22 & 0.22$\pm$0.16 \\
\tableline
\textbf{Dusty SNe} & \textbf{2.02$\pm$0.44} & \textbf{8.34$\pm$0.28} & \textbf{-2.02$\pm$1.01} & \textbf{0.27$\pm$0.21} \\
\textbf{H-rich} & 1.70$\pm$0.37 & 8.32$\pm$0.26 & -1.88$\pm$0.87 & 0.25$\pm$0.19 \\
\textbf{H-poor} & 2.05$\pm$0.59 & 8.47$\pm$0.39 & -1.86$\pm$1.26 & 0.39$\pm$0.27 \\ 
\textbf{IIn} & 1.77$\pm$0.64 &  8.35$\pm$0.27 & -3.58$\pm$0.40 & 0.18$\pm$0.06 \\
\tableline
\enddata
\tablecomments{The environmental properties of dusty SNe in bold are presented in terms of the three typical SN types (H-rich, H-poor, and type IIn), respectively. The values of dusty SNe in the three typical SN types are corrected based on the new SN type classification.} 
\end{deluxetable*}


\section{Environmental Properties} \label{sec:result}

In this section, we analyze the local environment of dusty SNe in our sample and compare 
the properties between the different CCSN types. The median and mean values of the environmental properties of dusty SNe are listed in Table \ref{tab:median4types}. We note a clear difference in H$\alpha$ equivalent width, metallicity, star formation rate and Extinction between dusty and typical SN types, which would be discussed more in detail in the following. \\

\subsection{H$\alpha$ equivalent width} \label{subsec:EW_Ha}

\begin{figure*}[!ht]
\centering
\includegraphics[width=0.8\columnwidth]{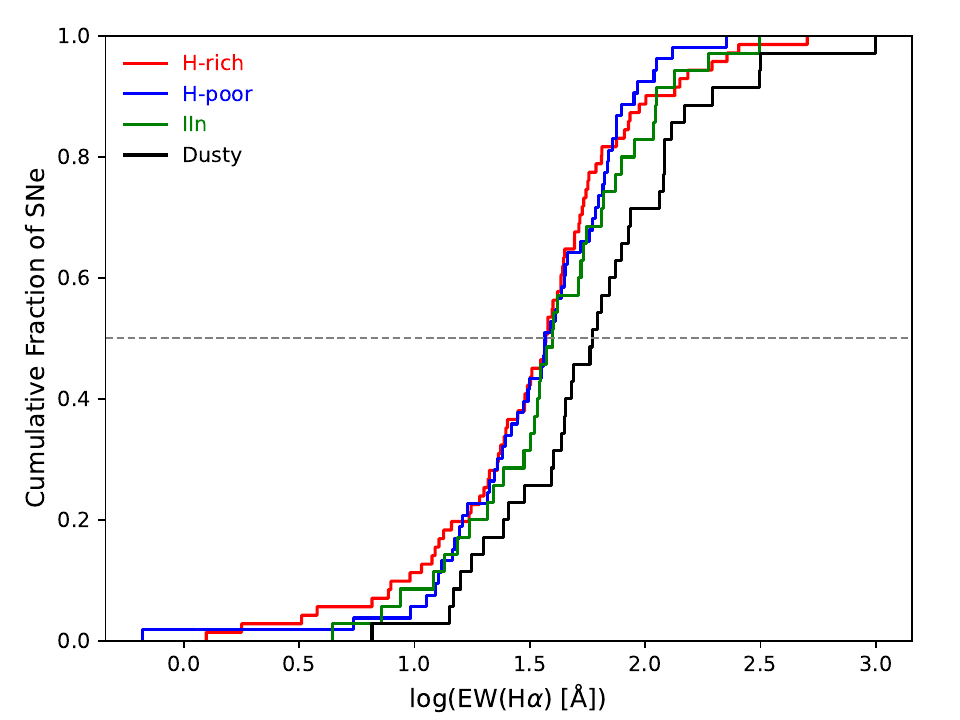}
\includegraphics[width=0.8\columnwidth]{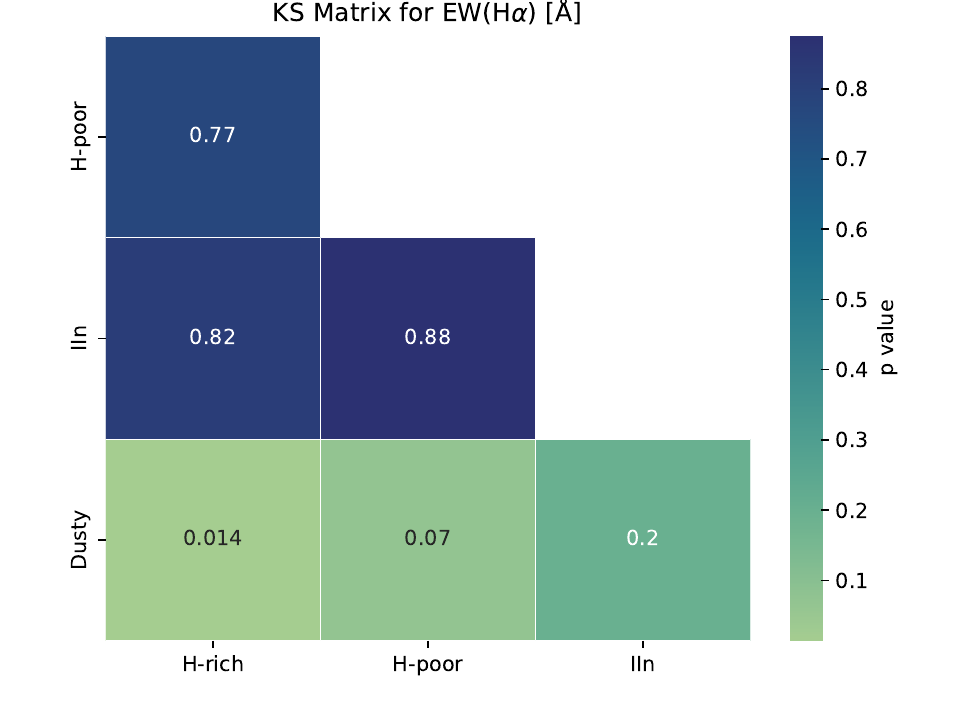}
\caption{Left: normalized cumulative distributions of H$\alpha$ equivalent width of SN hosted HII regions in four types: H-rich in blue, H-poor in red, IIn in green and dusty SNe in black. A dotted horizontal line at 0.5 fraction represents the median value of the distributions. Right: KS matrix for each combination of different SN types. 
\label{fig:hist_EW}}
\end{figure*}

Figure \ref{fig:hist_EW} shows the cumulative distributions of EW(H$\alpha$) of the SN-hosted HII regions for different CCSN types, and the KS test for each combination of them. We note that dusty SNe as a whole has a relatively higher EW(H$\alpha$) compared to the typical SN types which share a similar distribution in EW(H$\alpha$), which indicates a connection of dusty SNe with younger stellar populations. The differences of the distribution for each SN type combinations are also presented quantitatively in the KS matrix, and the $p$-value of the tests express the statistical significance of two distributions being drawn from the same parent sample. We choose a confidence level of 95\%, that is, we consider that a $p-$value $\leq$ 0.05 as indicating a statistically significant difference between these two distributions. Therefore, dusty SNe show statistically significant differences with hydrogen-rich SNe in terms of EW(H$\alpha$), although more than 60\% of dusty SNe are hydrogen-rich ones. 

\begin{figure*}[ht!]
\centering
\includegraphics[width=0.8\columnwidth]{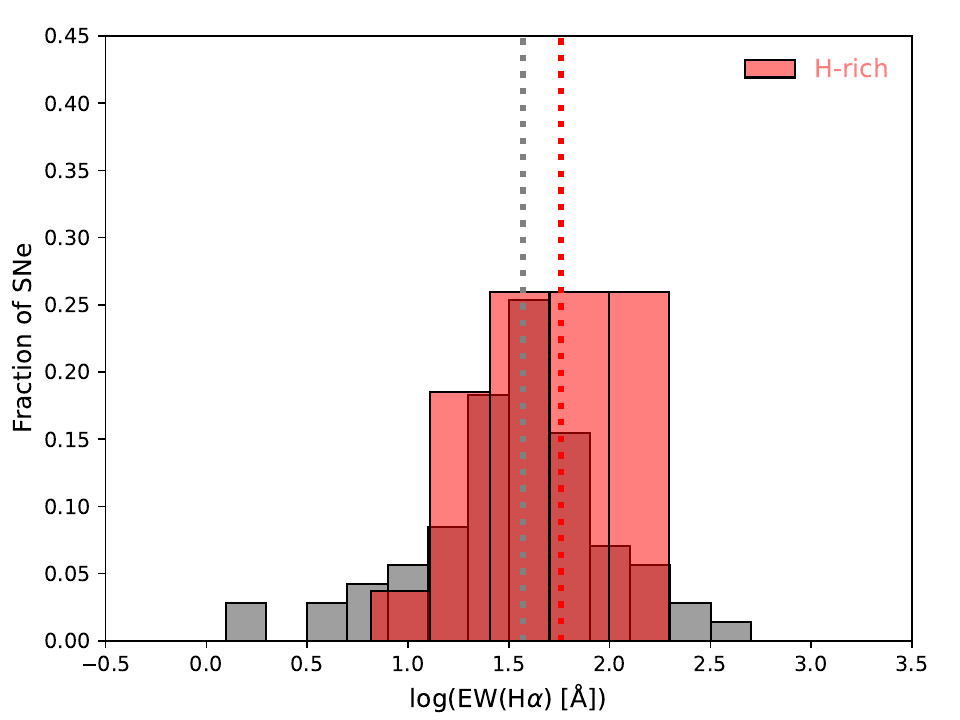}
\includegraphics[width=0.8\columnwidth]{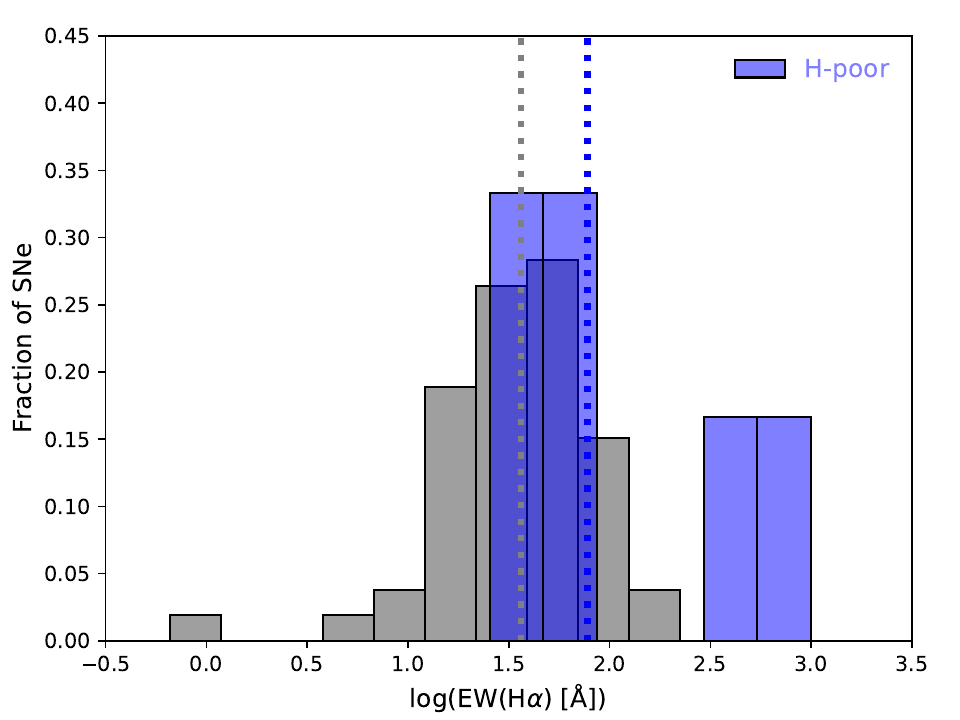}\\ 
\includegraphics[width=0.8\columnwidth]{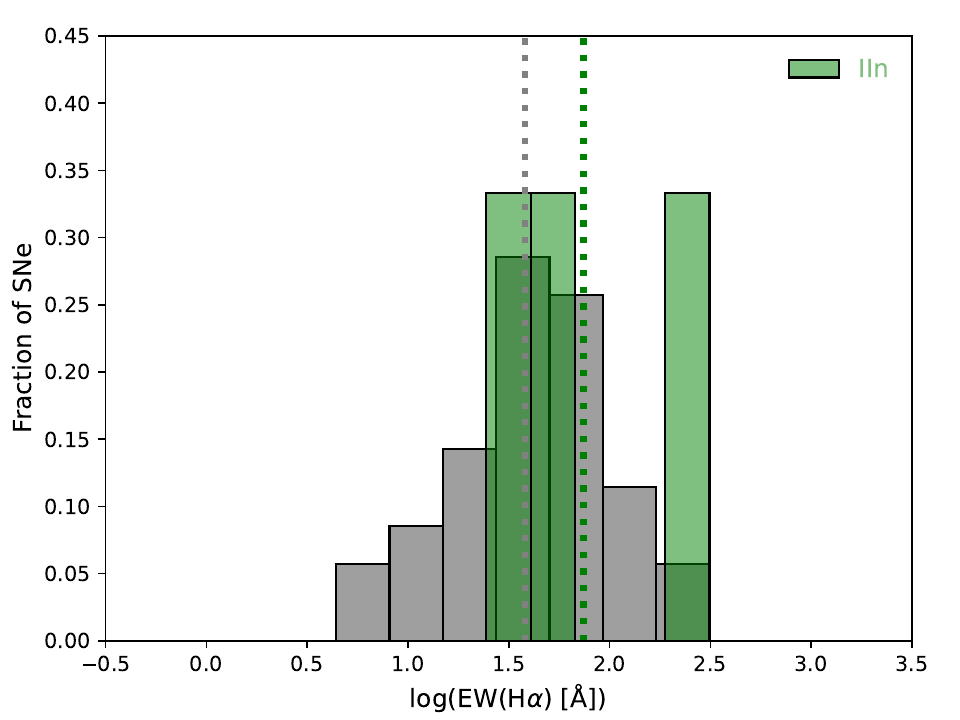}
\includegraphics[width=0.8\columnwidth]{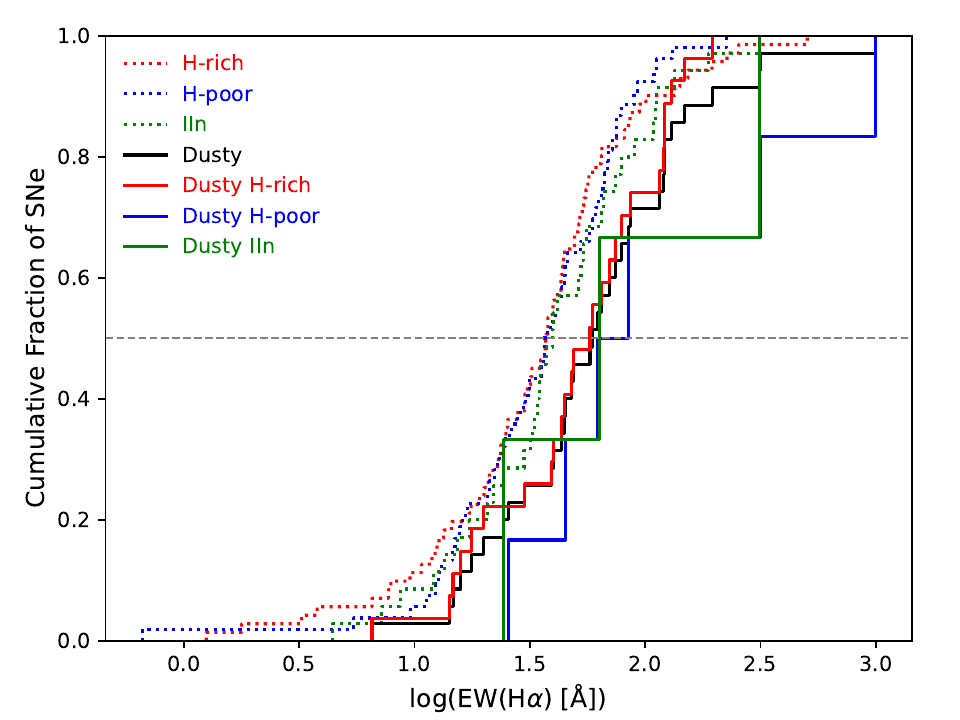}
\caption{
Histograms of H$\alpha$ equivalent width of dusty SNe in three types: H-rich in blue (the top left), H-poor in red (the top right), and IIn in green ( the bottom left) separately, compared with typical SN types in gray. The median values of different SNe types are marked by the vertical lines. In the bottom right, the normalized cumulative distributions are compared between typical SNe in dotted lines and dusty SNe in solid lines. 
\label{fig:hist_EW_dusty}}
\end{figure*}

In Figure \ref{fig:hist_EW_dusty}, we further study the EW(H$\alpha$) of dusty SN hosts in different types, H-rich, H-poor and IIn respectively. In general, dusty SN hosts prefer the location of higher H$\alpha$ equivalent width in all subtypes, and this preference increases from H-rich, type IIn and to H-poor. Considering the meaning of EW(H$\alpha$) as stellar age indicator, the increasing sequence from H-rich, type IIn to H-poor indicates the decrease of average age, which is consistent with the predicted stellar age of the corresponding SN progenitors \citep{Smartt2009,Smartt2015}. \\

\subsection{Star formation rate} \label{subsec:SFR} 

\begin{figure*}[ht!]
\centering
\includegraphics[width=0.8\columnwidth]{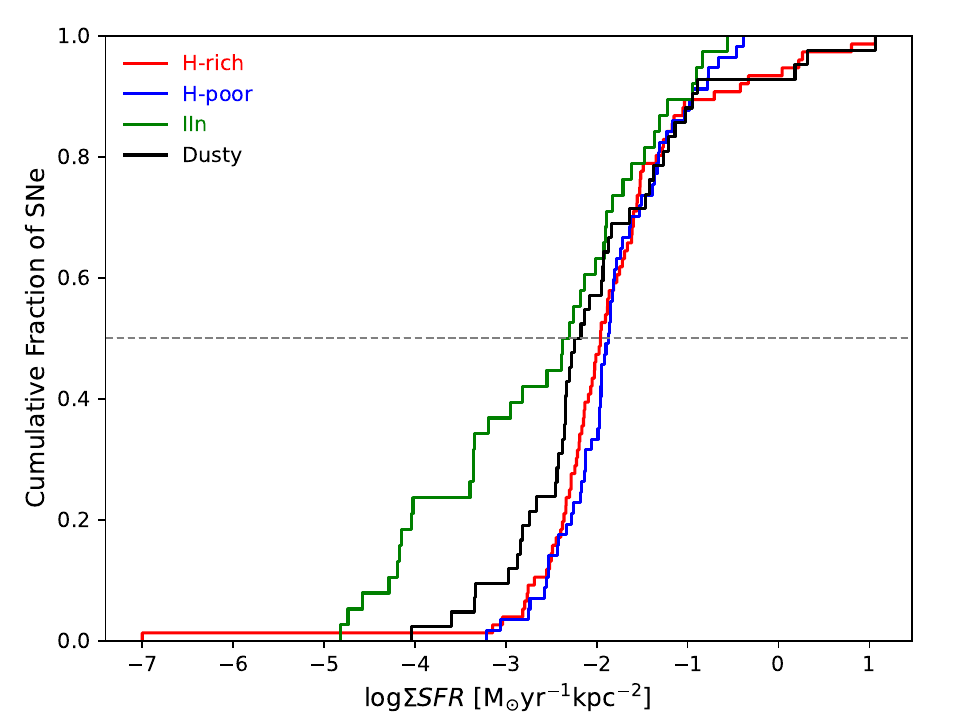}
\includegraphics[width=0.8\columnwidth]{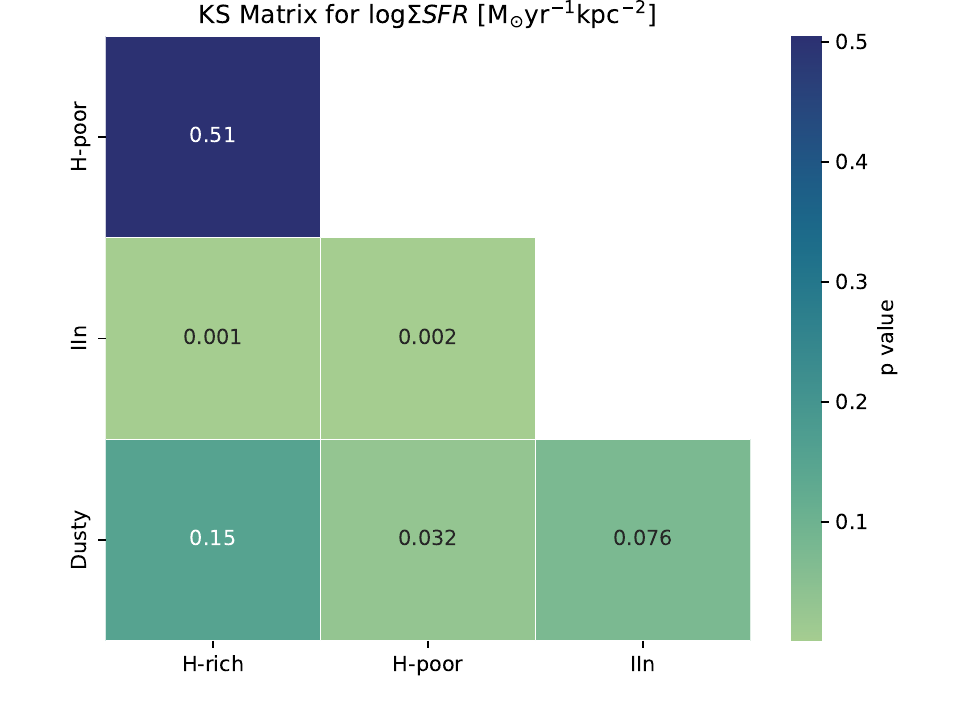}\\
\caption{Left: normalized cumulative distributions of star formation rate intensity at SN sites are compared between dusty SNe in black and three typical SN types with H-rich in blue, H-poor in red, and IIn in green. A dotted horizontal line at 0.5 fraction represents the median value of the distributions. Right: KS statistic matrix for each combination of SN types.
\label{fig:hist_SFR}}
\end{figure*}

\begin{figure*}[ht!]
\centering
\includegraphics[width=0.8\columnwidth]{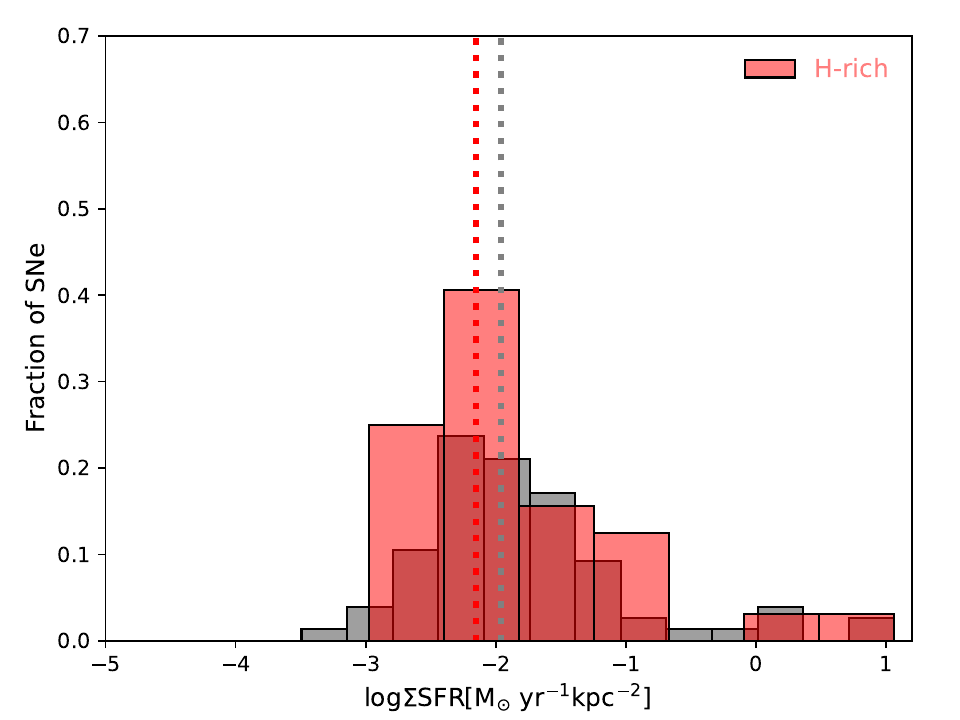}
\includegraphics[width=0.8\columnwidth]{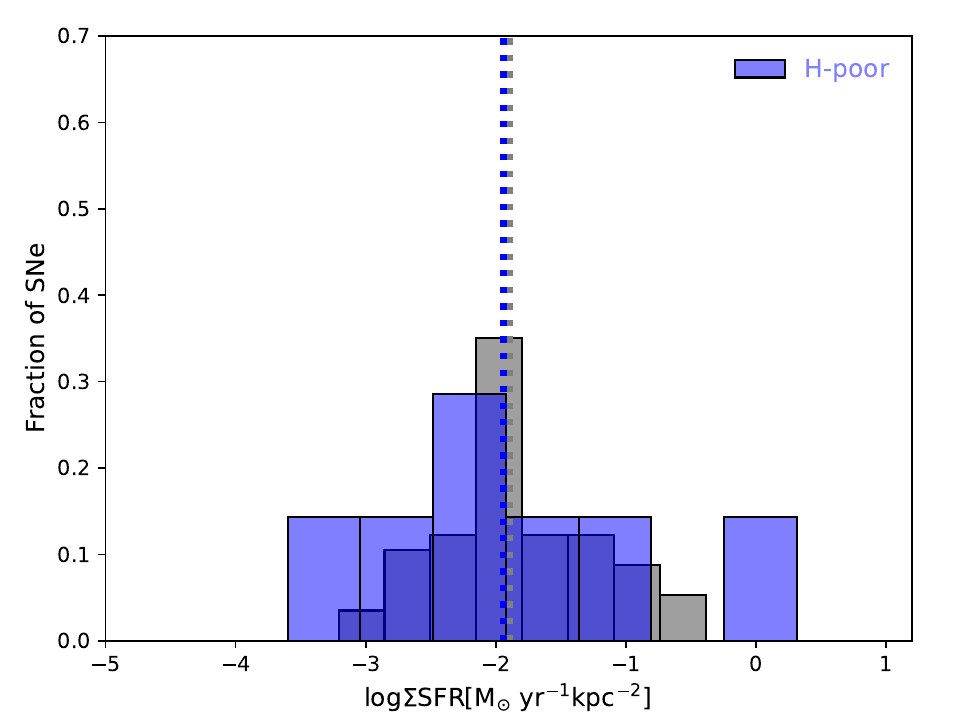}\\
\includegraphics[width=0.8\columnwidth]{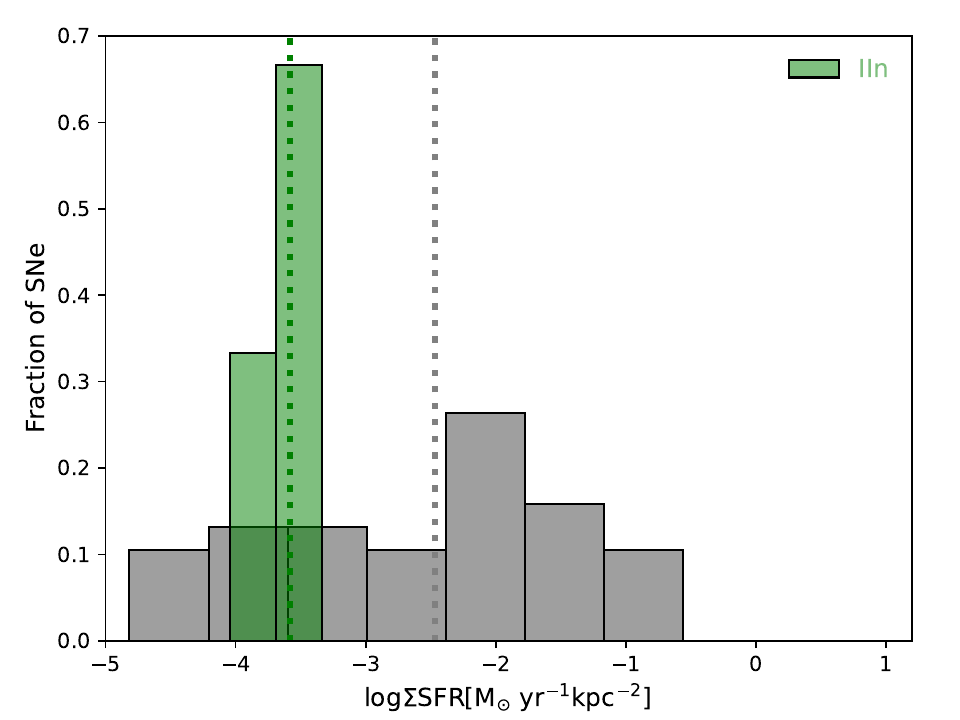}
\includegraphics[width=0.8\columnwidth]{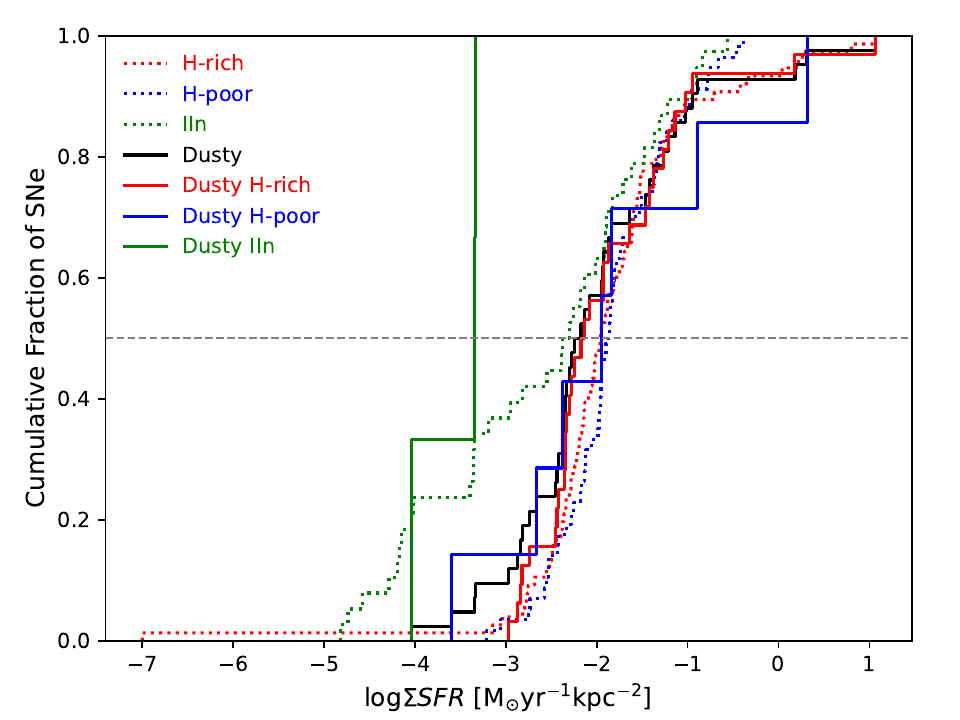}
\caption{
Histograms of star formation rate (SFR) of dusty SNe in three types: H-rich in blue (the top left), H-poor in red (the top right), and IIn in green ( the bottom left) separately, compared with typical SN types in gray. The median values of different SNe types are marked by the vertical lines. In the bottom right, the normalized cumulative distributions are compared between typical SNe in dotted lines and dusty SNe in solid lines. \\
\label{fig:hist_SFR_dusty}}
\end{figure*} 

Figure \ref{fig:hist_SFR} presents the cumulative distributions of SFR intensity in the SN host HII regions for different SN subtypes and the KS test of each combination. H-rich and H-poor SNe show a similar distribution of star formation rate and tend to occur at higher SFR regions, followed by dusty SNe and type IIn SNe. There is an increasing deviation for the SFR distributions of different types as the host SFR decreases, in particular for type IIn SNe. The significant difference in SFR between type IIn with H-rich and H-poor SNe are also presented in the KS test matrix of $p$-value $<$ 0.05. The dusty SNe again statistically differs with H-poor SNe in terms of star formation rate with a relatively lower SFR, but do not have a statistically significant difference with type IIn and H-rich SNe any more as shown in the KS matrix. 

In Figure \ref{fig:hist_SFR_dusty}, the SFR distribution of dusty SN hosts in three typical SN types are presented. We note that H-rich and H-poor dusty SNe span a similar distribution with the corresponding typical SN types. On the other hand, the type IIn dusty SNe are more likely discovered at lower star formation rate regions compared to the typical type. In general, H-rich and H-poor SNe are more correlated with ongoing massive star formation than IIn SNe. Type IIn SNe contribute to more dust at the region of lower SFR than other CCSN types. Considering that the sample of type IIn dusty SNe is small, it is difficult to draw any definitive conclusions about the overall trend. 

We also note that there are inconsistencies between the EW(H$\alpha$) and SFR distributions, and it would be necessary to investigate the physical processes described by these two parameters. The star formation rate intensity log$\Sigma_{\rm SFR}$ here is measured from the extinction-corrected H${\rm \alpha}$ flux contributed by young massive stars, tracing the ongoing SFR. EW(H${\rm \alpha}$) is measured by the strength of the line relative to the continuum, which in fact can be an indicator of the strength of the ongoing SFR compared with the past SFR, which reduces with time if no new stars are created. However, considering the size of HII region segregation in IFS can be as large as hundreds of parsec in radius, the continuous star formation bursts in the same site are more reasonable, compared to a single instantaneous star formation history. Therefore, the contamination from older stellar populations together with effect of photo-leakage can bring uncertainties to the estimation of stellar population age \citep{Sun2021}. \\

\subsection{Oxygen abundance} \label{subsec:OH}

\begin{figure*}[ht!]
\centering
\includegraphics[width=0.8\columnwidth]{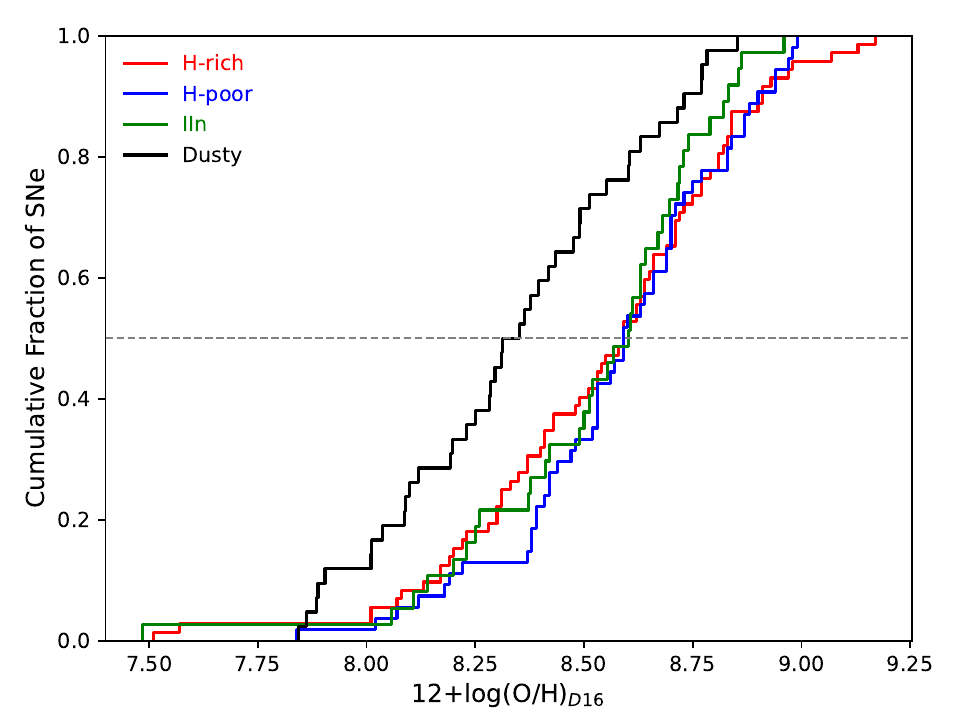}
\includegraphics[width=0.8\columnwidth]{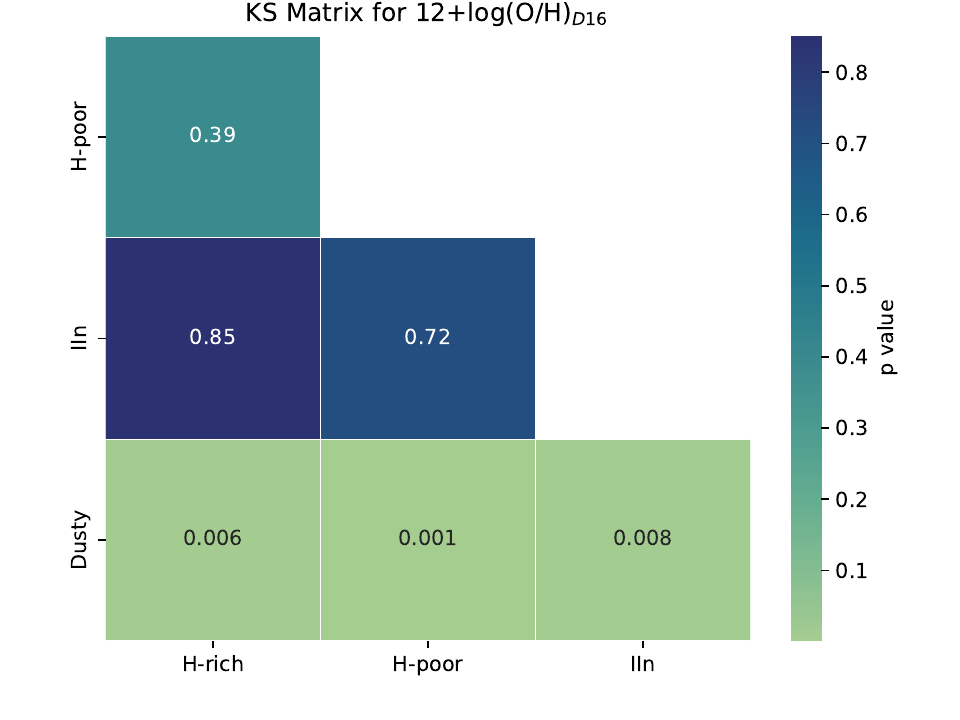}\\
\caption{Left: normalized cumulative distributions of oxygen abundance (12+log(O/H)$_{\rm D16}$) at SN sites are compared between dusty SNe in black and three typical SN types with H-rich in blue, H-poor in red, and IIn in green. A dotted horizontal line at 0.5 fraction represents the median value of the distributions. Right: KS statistic matrix for each combination of SN types. 
\label{fig:hist_OH}}
\end{figure*}

Metallicity plays a crucial role in the evolution of massive stars and their final CCSN explosions. Oxygen is the most abundant metal in the gas phase and exhibits very strong nebular lines at optical wavelengths, so that is usually chosen as a metallicity indicator. The cumulative distributions of oxygen abundance of the SN host HII regions for different SN subtypes and the KS test for each combination, are presented in Figure \ref{fig:hist_OH}. The dusty SN hosts are located at lower metallicity regions with the mean and median values of 12+log(O/H)$_{\rm D16} \sim $ 8.33 dex, which is more than 0.2 dex lower than those of typical CCSN host regions as shown in Table \ref{tab:sndata-photo}. By contrast, the three typical CCSN hosts span a similar distribution in a range from 12+log(O/H)$_{\rm D16} \sim $ 8.0 to 8.8 dex. Additionally, the KS test with $p$-value $<$ 0.05 also suggests that there is statistically significant difference between dusty SNe and the typical CCSN types in oxygen abundance. 

\begin{figure*}[ht!]
\centering
\includegraphics[width=0.8\columnwidth]{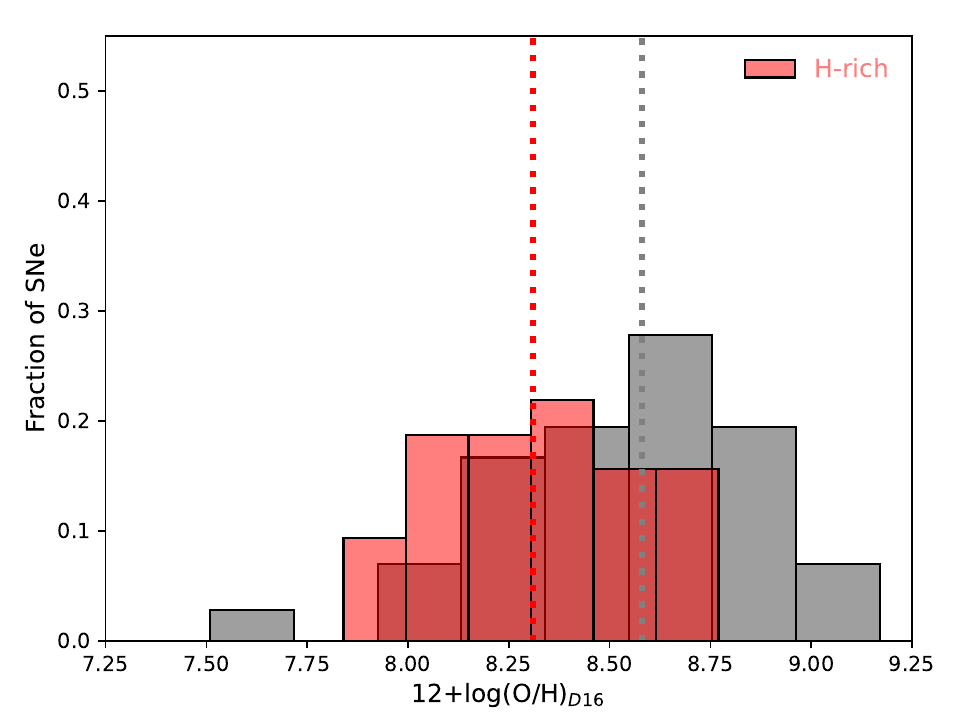}
\includegraphics[width=0.8\columnwidth]{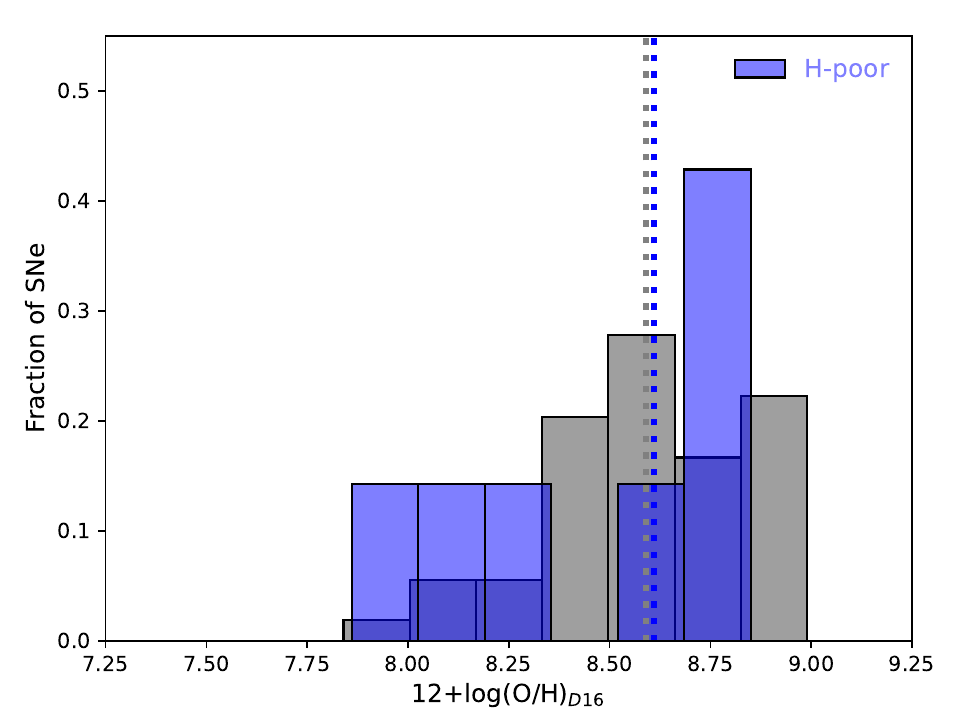}\\
\includegraphics[width=0.8\columnwidth]{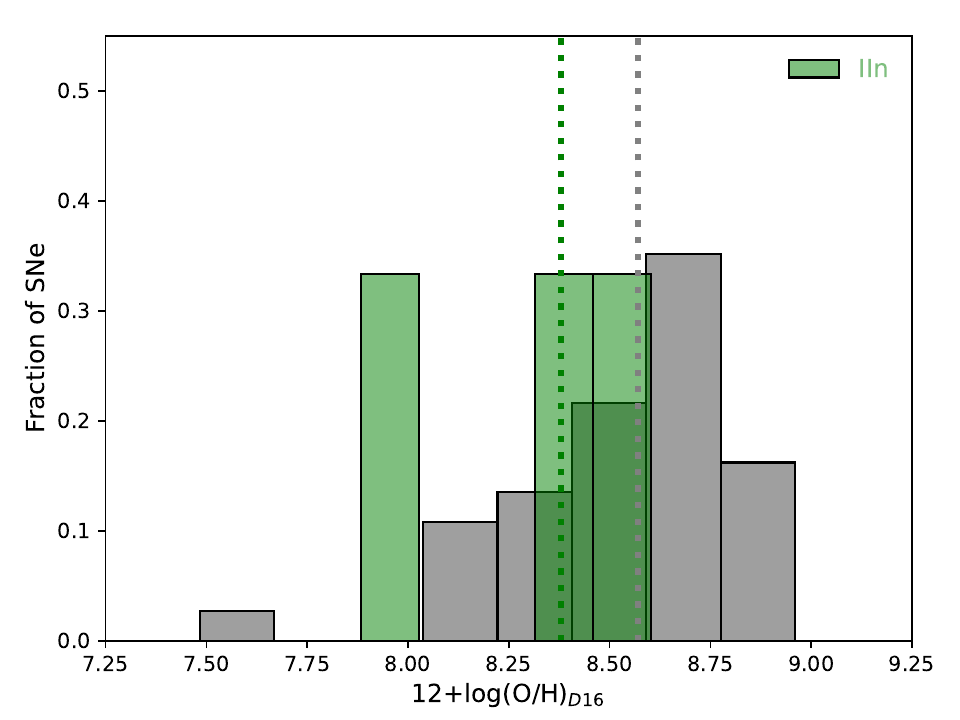}
\includegraphics[width=0.8\columnwidth]{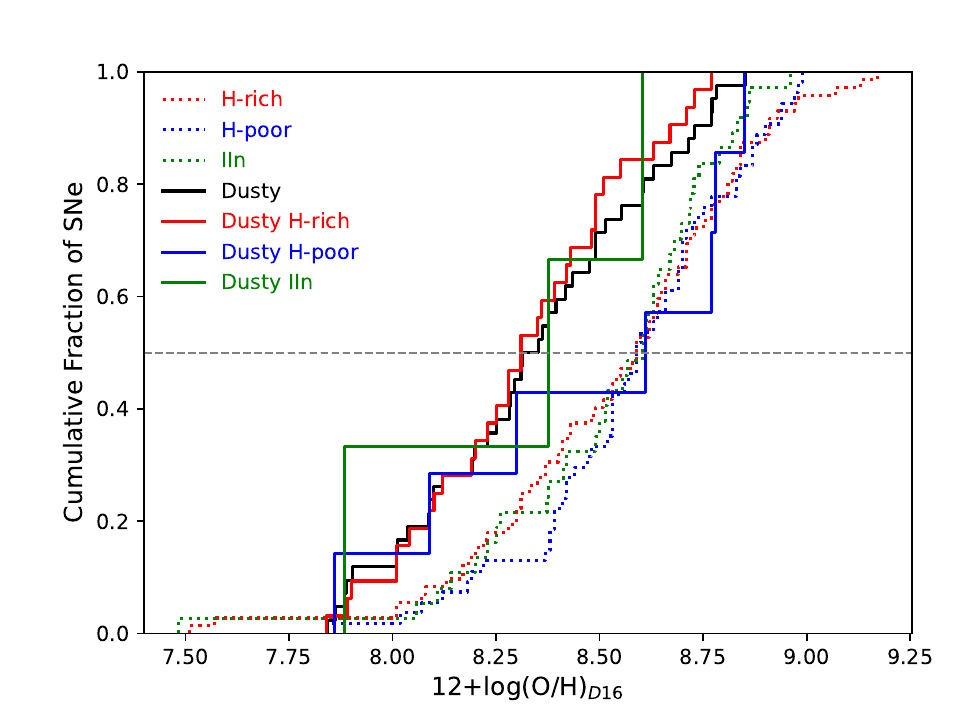}\\
\caption{Histograms of oxygen abundance (12+log(O/H)$_{\rm D16}$) of dusty SNe in three types: H-rich in blue (the top left), H-poor in red (the top right), and IIn in green ( the bottom left) separately, compared with typical SN types in gray. The median values of different SNe types are marked by the vertical lines. In the bottom right, the normalized cumulative distributions are compared between typical SNe in dotted lines and dusty SNe in solid lines. \\ 
\label{fig:hist_OH_dusty}}
\end{figure*}

\begin{figure*}[ht!]
\centering
\includegraphics[width=0.8\columnwidth]{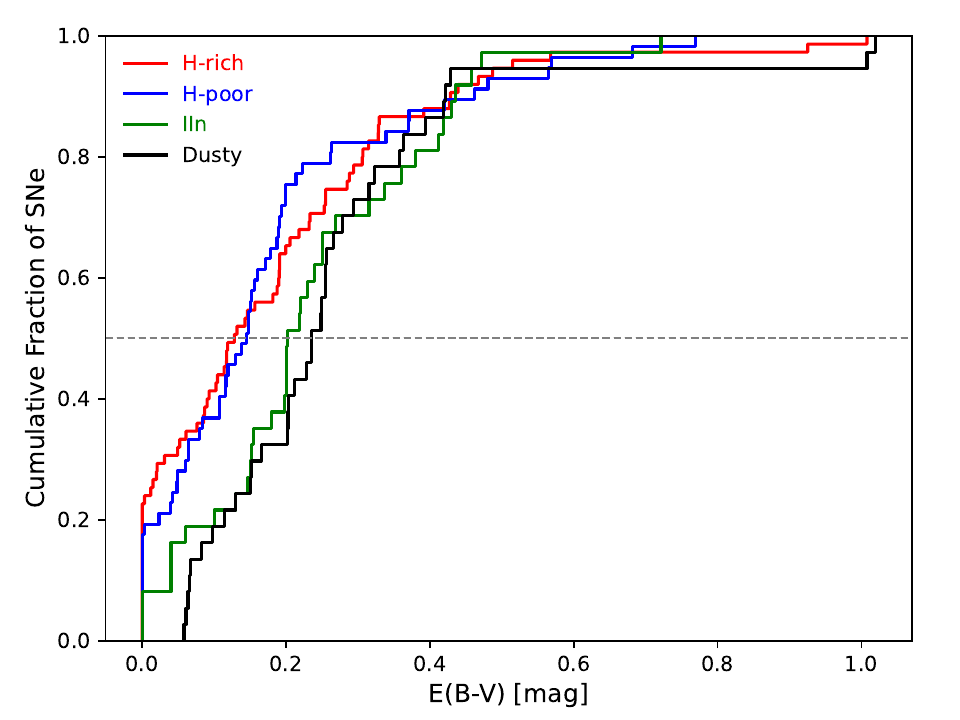}
\includegraphics[width=0.8\columnwidth]{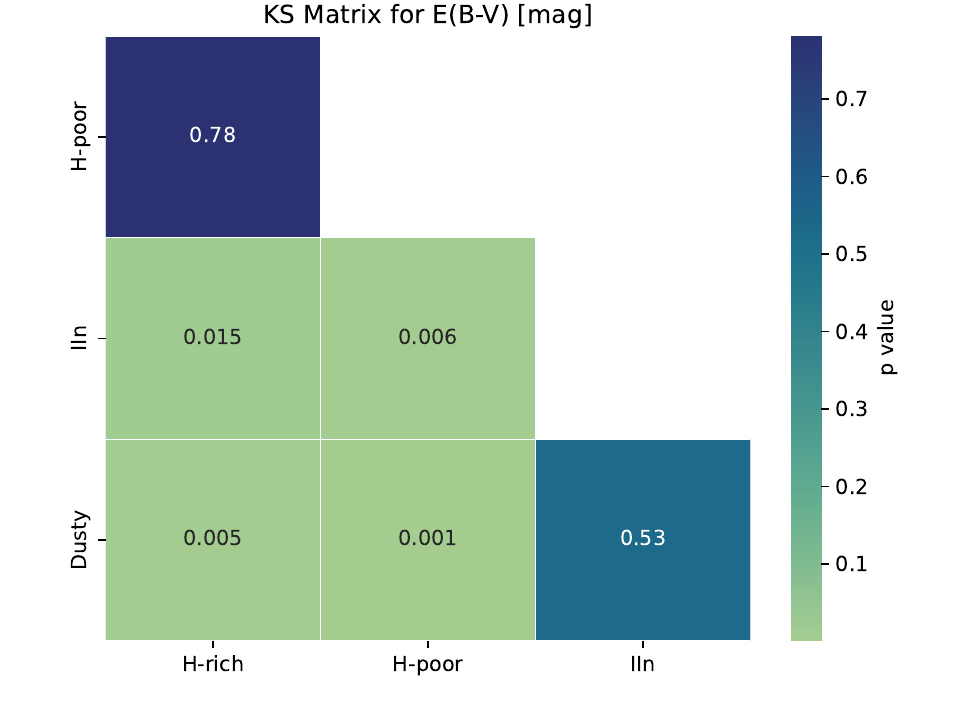}\\
\caption{Left: normalized cumulative distributions of E(B-V) at SN sites are compared between dusty SNe in black and three typical SN types with H-rich in blue, H-poor in red, and IIn in green. A dotted horizontal line at 0.5 fraction represent the median value of the distributions. Right: KS statistic matrix for each combination of SN types.
\label{fig:hist_EBV}}
\end{figure*}

\begin{figure*}[ht!]
\centering
\includegraphics[width=0.8\columnwidth]{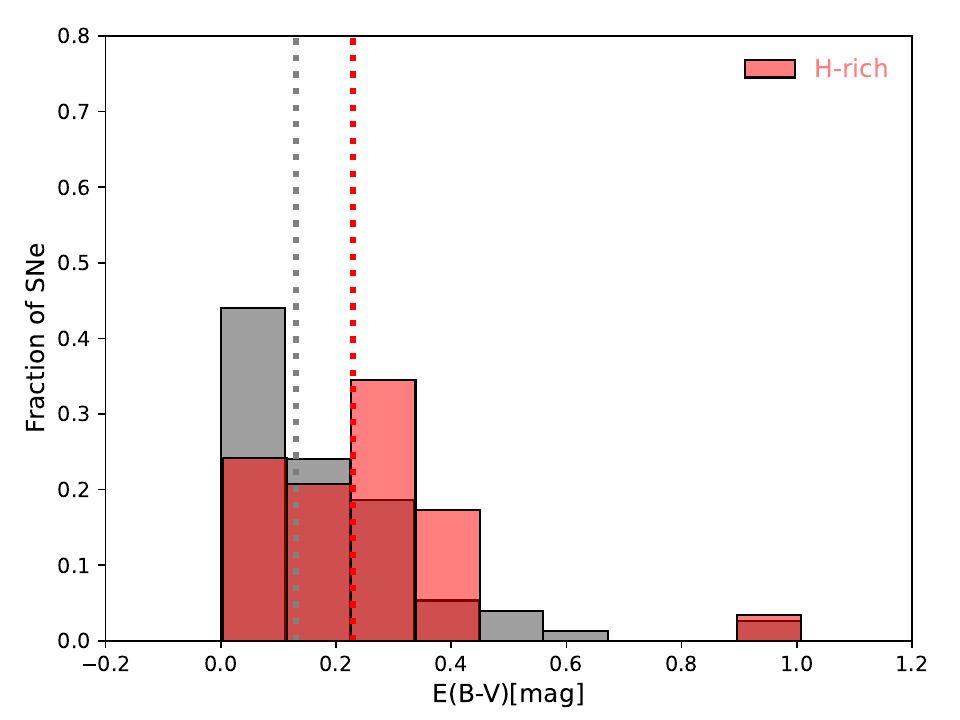}
\includegraphics[width=0.8\columnwidth]{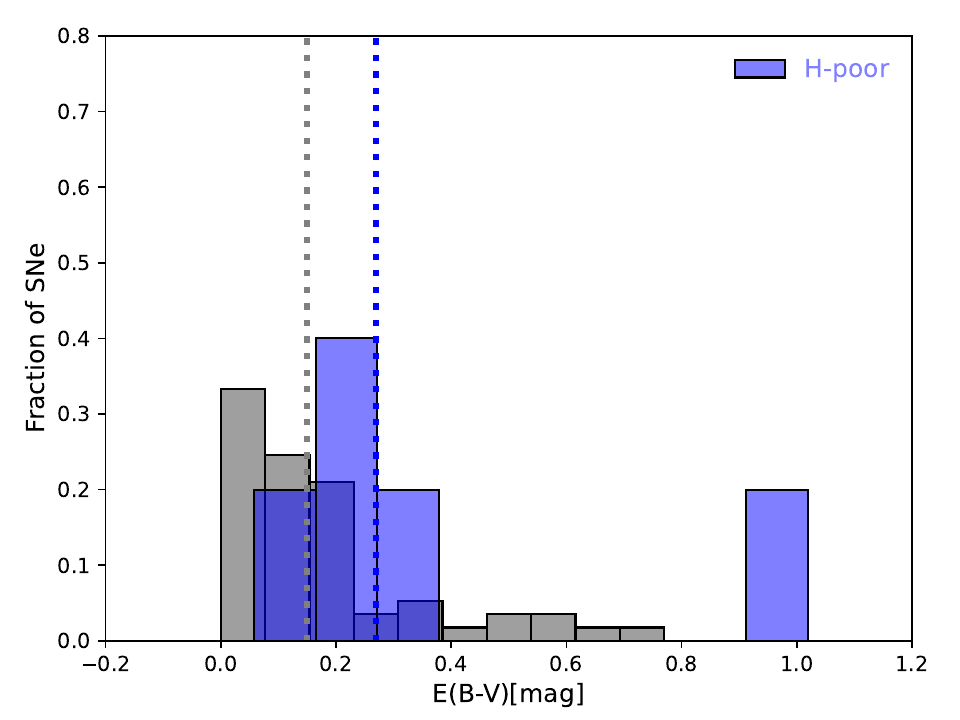}\\
\includegraphics[width=0.8\columnwidth]{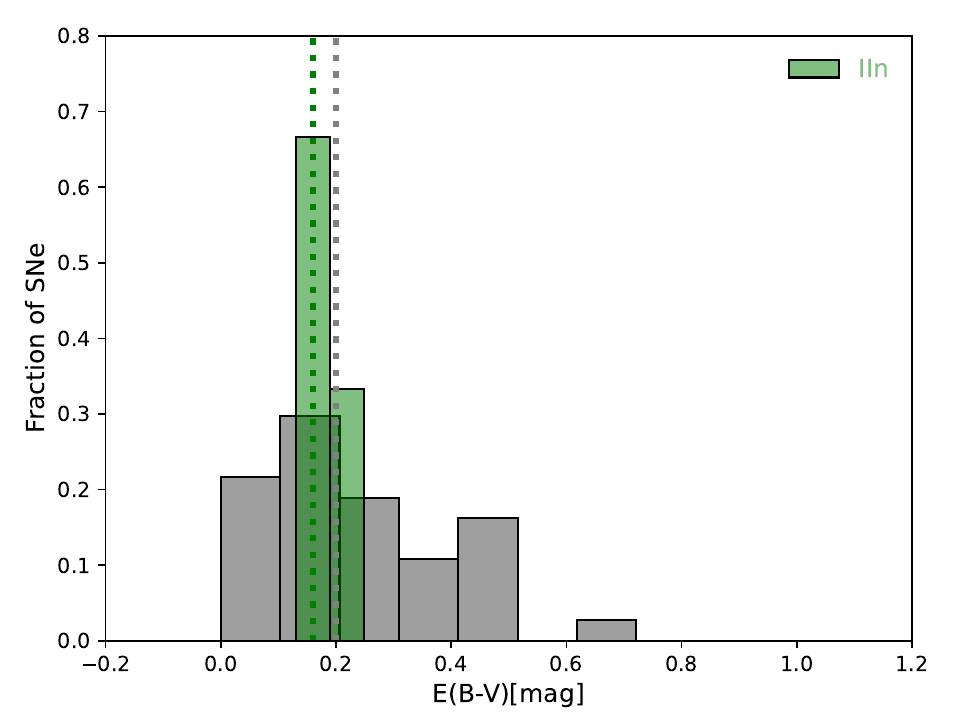}
\includegraphics[width=0.8\columnwidth]{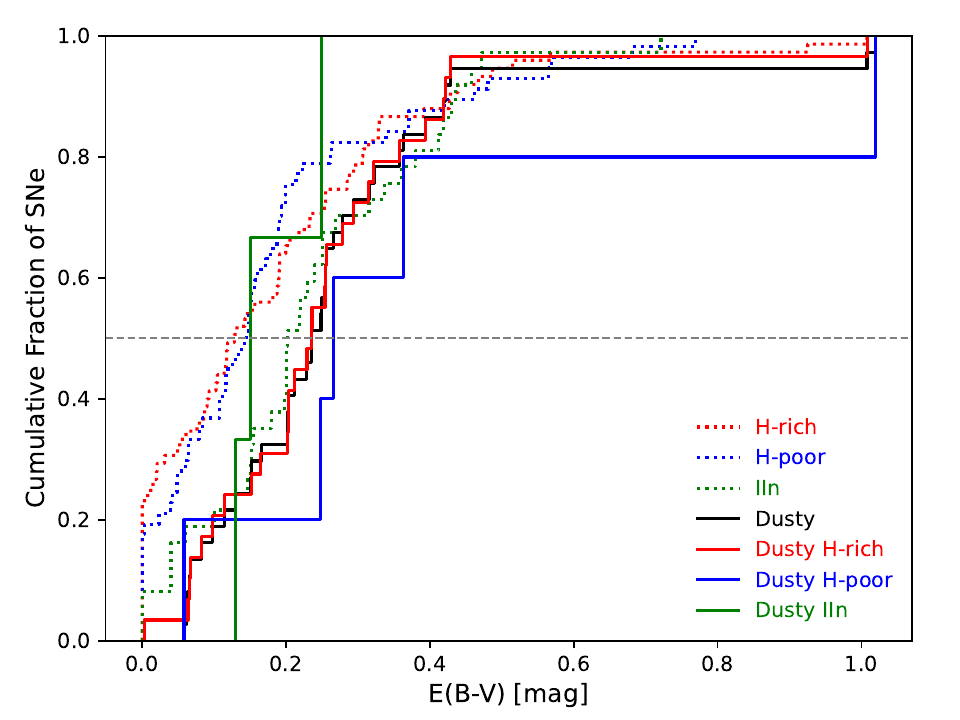}
\caption{
Histograms of E(B-V) of dusty SNe in three types: H-rich in blue (the top left), H-poor in red (the top right), and IIn in green ( the bottom left) separately, compared with typical SN types in gray. The median values of different SNe types are marked by the vertical lines. In the bottom right, the normalized cumulative distributions are compared between typical SNe in dotted lines and dusty SNe in solid lines. 
\label{fig:hist_EBV_dusty}}
\end{figure*}

In Figure \ref{fig:hist_OH_dusty}, we further compare the oxygen abundance of dusty SN hosts in different types, H-rich, H-poor and IIn respectively. We find that compared to the typical types, the distributions of H-rich and type IIn dusty SNe are shifted to lower metallicity regions, while H-poor dusty SNe span a similar range. As a consequence, a clear sequence in terms of average metallicity from low to high emerges, H-rich, type IIn and H-poor. This may indicates that H-poor dusty SNe are less sensitive with metallicity, and turn out to be the main dust contributor in higher metallicity regions. Considering the effect of metallicity on mass loss history of SN progenitors which construct the CSM surrounding SNe, the difference in metallicity of different dusty SN types may reflect the diversity in the evolution of the corresponding SN progenitors. H-rich SNe are usually thought to be arise from massive stars experiencing a single-star evolution, whose mass loss are metallicity dependent driven by stellar wind. By contrast, H-poor SN progenitors are more dominated by lower-mass stars in binary systems with massloss through binary interactions, which is less dependent on metallicity \citep{Eldridge2017}. The progenitor populations of type IIn SNe can be a combination of both single and binary star system, so that its metallicity is in the middle between H-rich and H-poor SNe. \\\\

\subsection{Extinction} \label{subsec:EBV}

The extinction for SN host HII regions is also related to the extinction suffered by the SN themselves. We therefore compare the values of extinction between dusty SNe with the different CCSN types and analyze which subtype environment is more reddened, and therefore more affected by the presence of pre-existing dust. Figure \ref{fig:hist_EBV} presents the cumulative distributions for the SN host extinction in different SN types and reported the KS matrix for each combination of them. We note that dusty SNe have a similar distribution with the traditionally CSM-interaction SNe, type IIn, which are more likely located at higher extinction regions. On the other hand, H-rich and H-poor SNe also share a similar range in extinction but with lower values of E(B-V). Therefore, the SN dust formation may correlate to higher extinction host environments, where CSM interactions are more effective.

Then, the KS matrix with $p$-value $<$ 0.05 in the combination of type IIn and dusty with H-rich and H-poor SNe, suggests a statistically significant difference between these types. In addition, Figure \ref{fig:hist_EBV_dusty} presents the extinction difference of dusty SN host environment in different types. We note that most dusty SNe shift to higher extinction region, \text{red}{except for type IIn dusty SNe which host at lower extinction region.} Therefore, it seems to be critical for H-rich and H-poor SNe to explode in a more dusty environment to become an observed dusty SNe in mid-IR.  In fact, the mid-IR studies have revealed that a significant amount of newly-formed dust in SN ejecta and SN shocked regions is attributed to the observed luminous mid-IR emissions \citep{Fox2011,Szalai2013,Smith2017}. \\\\

\section{Environmental dependence of SN dust} \label{sec:dependence}


\begin{figure*}[ht!]
\centering
\includegraphics[width=0.8\columnwidth]{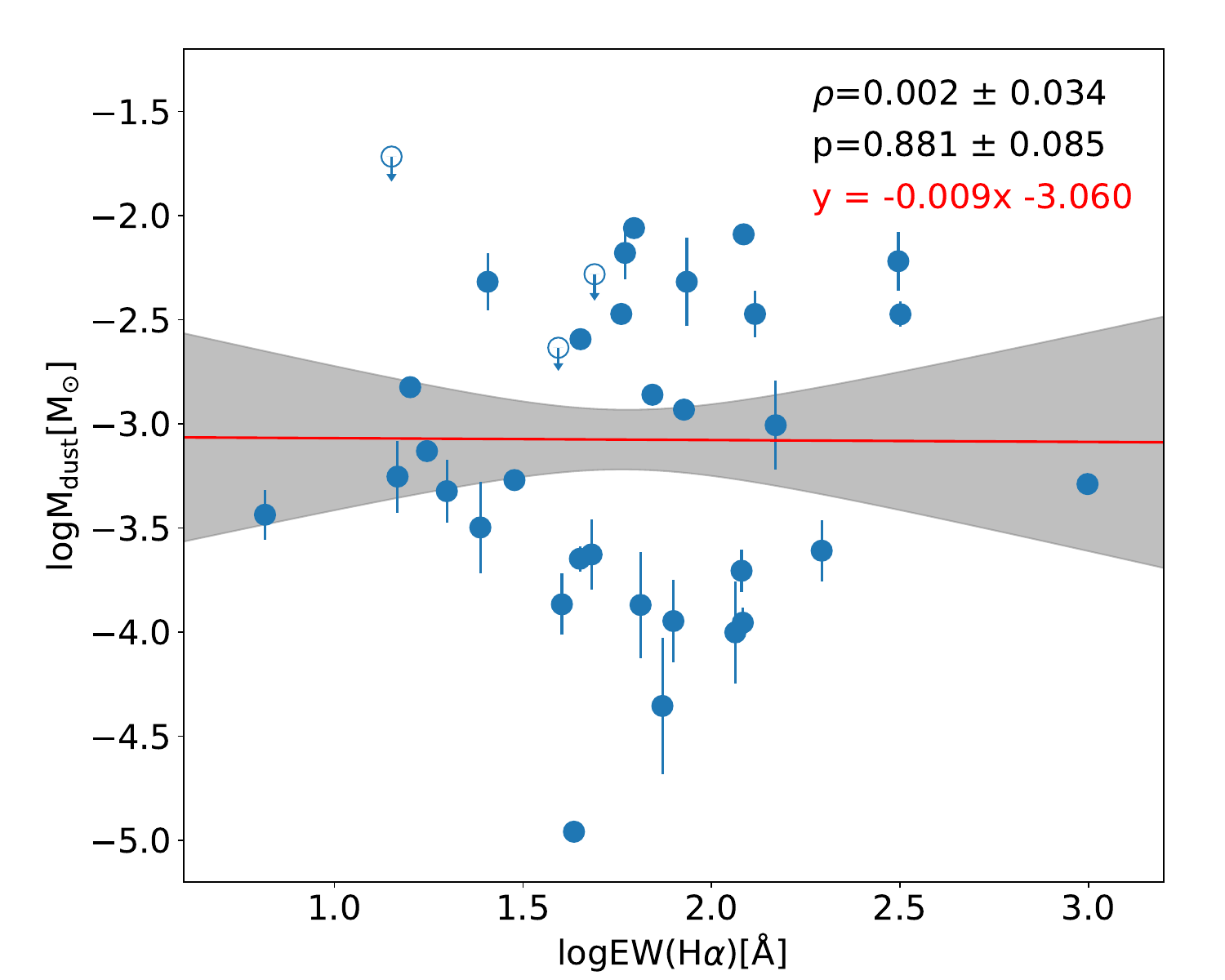}
\includegraphics[width=0.8\columnwidth]{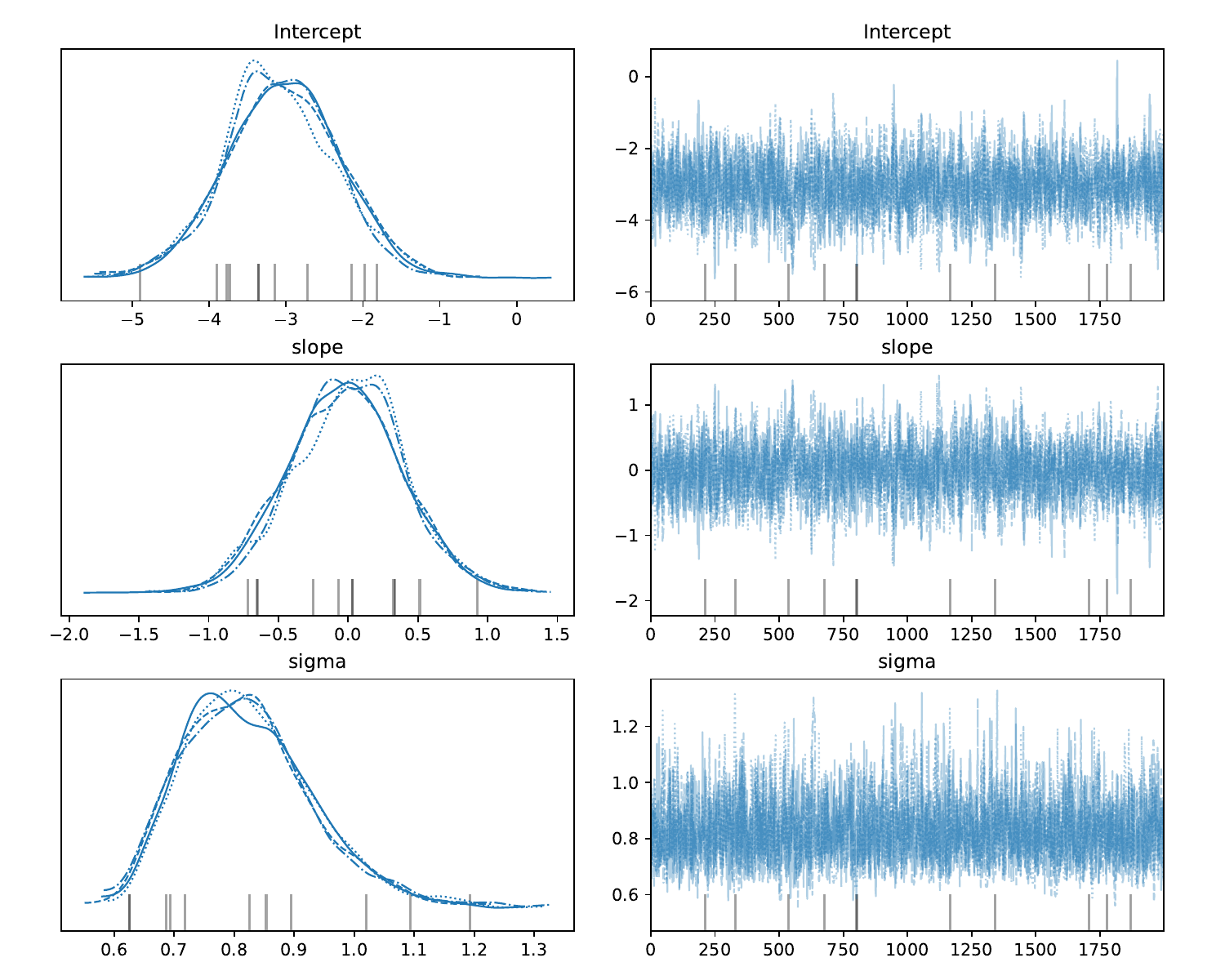}\\
\includegraphics[width=0.8\columnwidth]{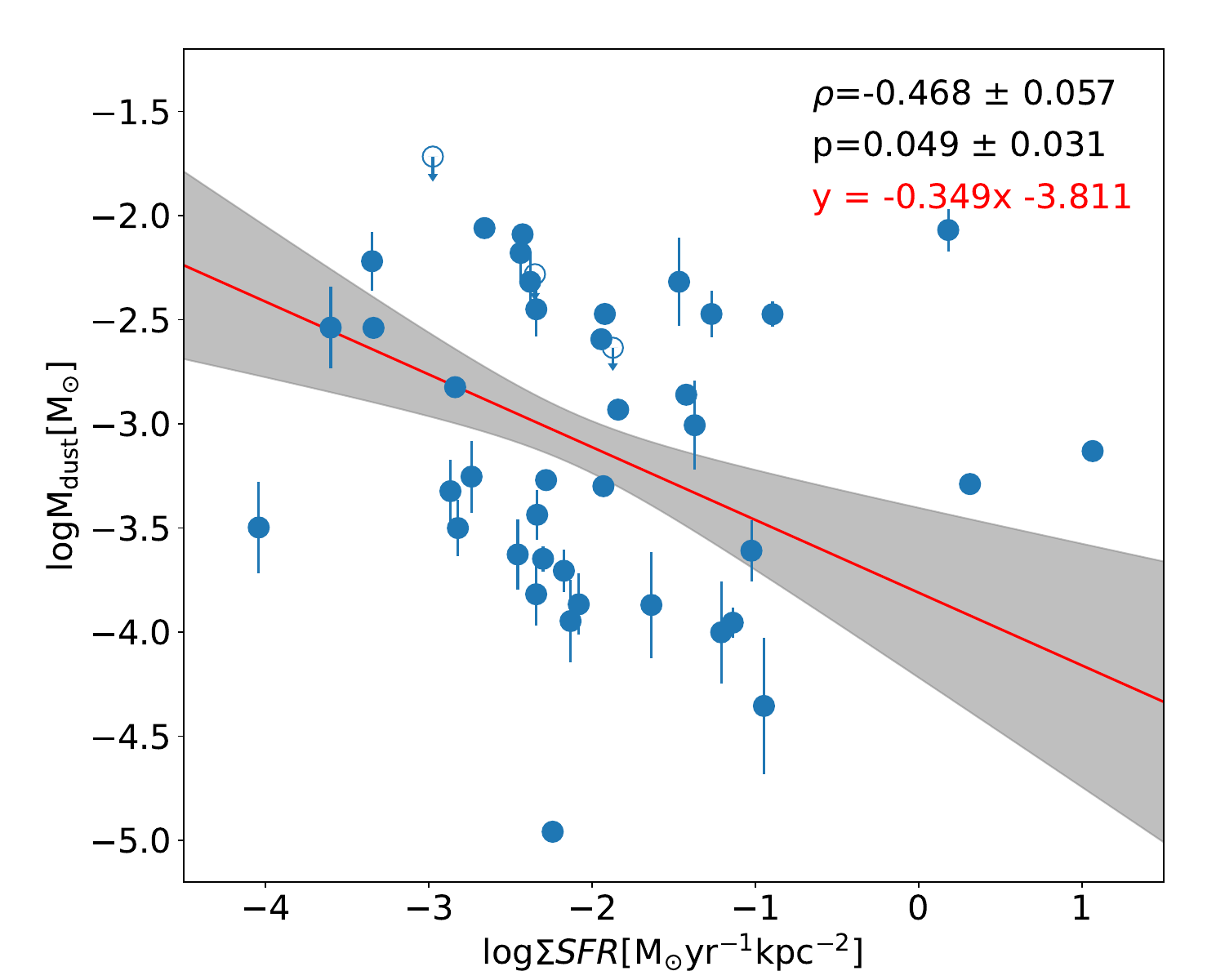}
\includegraphics[width=0.8\columnwidth]{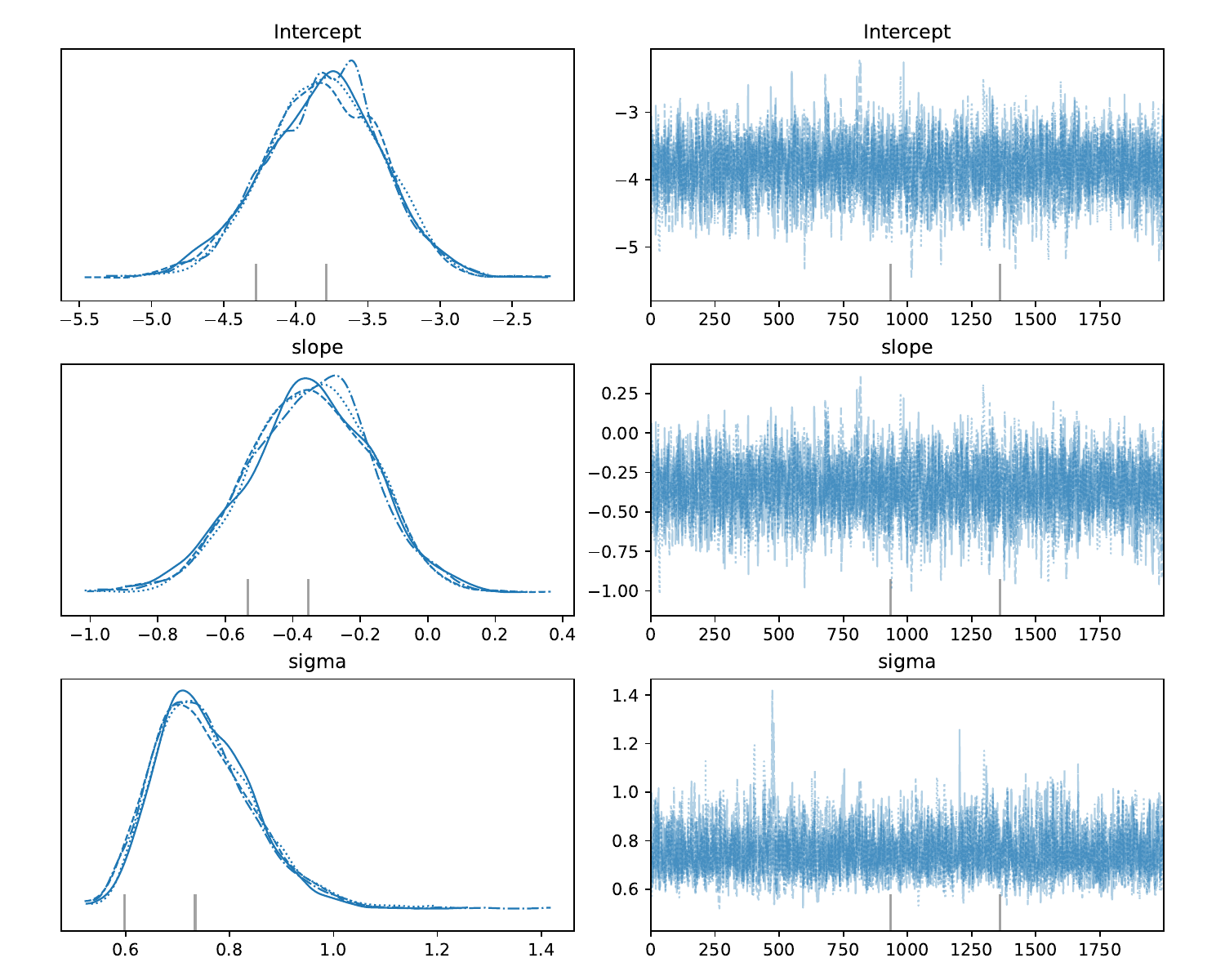}\\
\includegraphics[width=0.8\columnwidth]{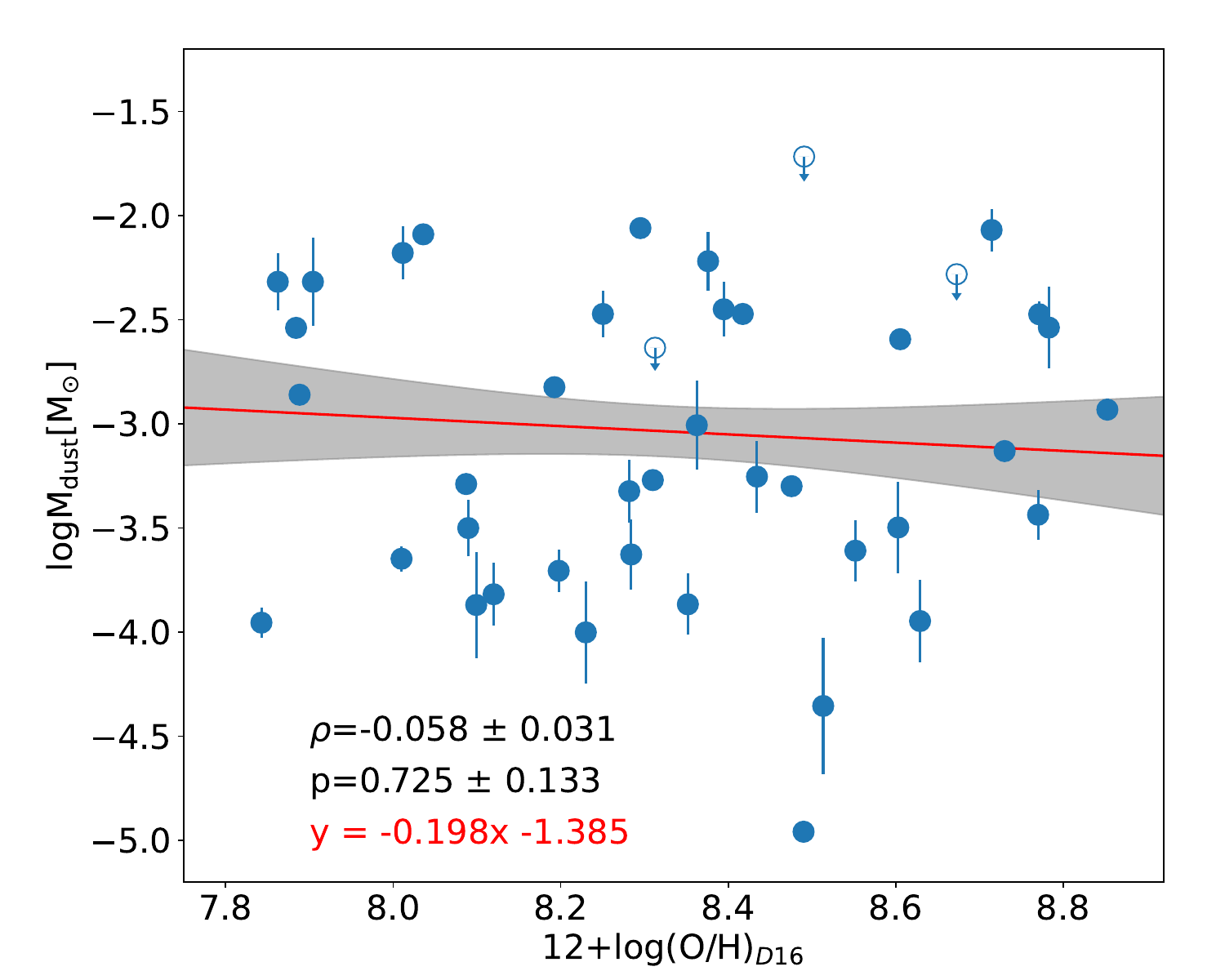}
\includegraphics[width=0.8\columnwidth]{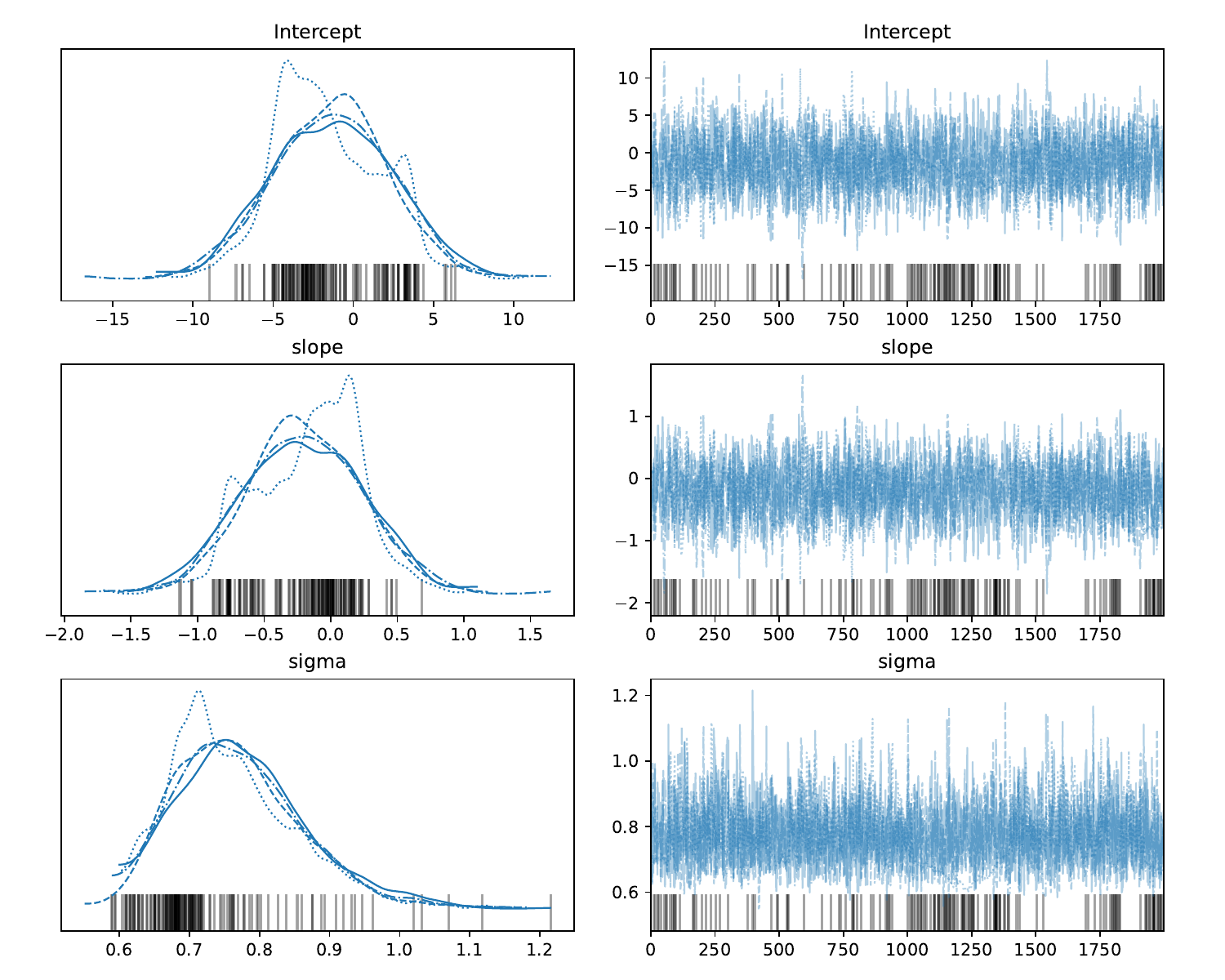}\\
\includegraphics[width=0.8\columnwidth]{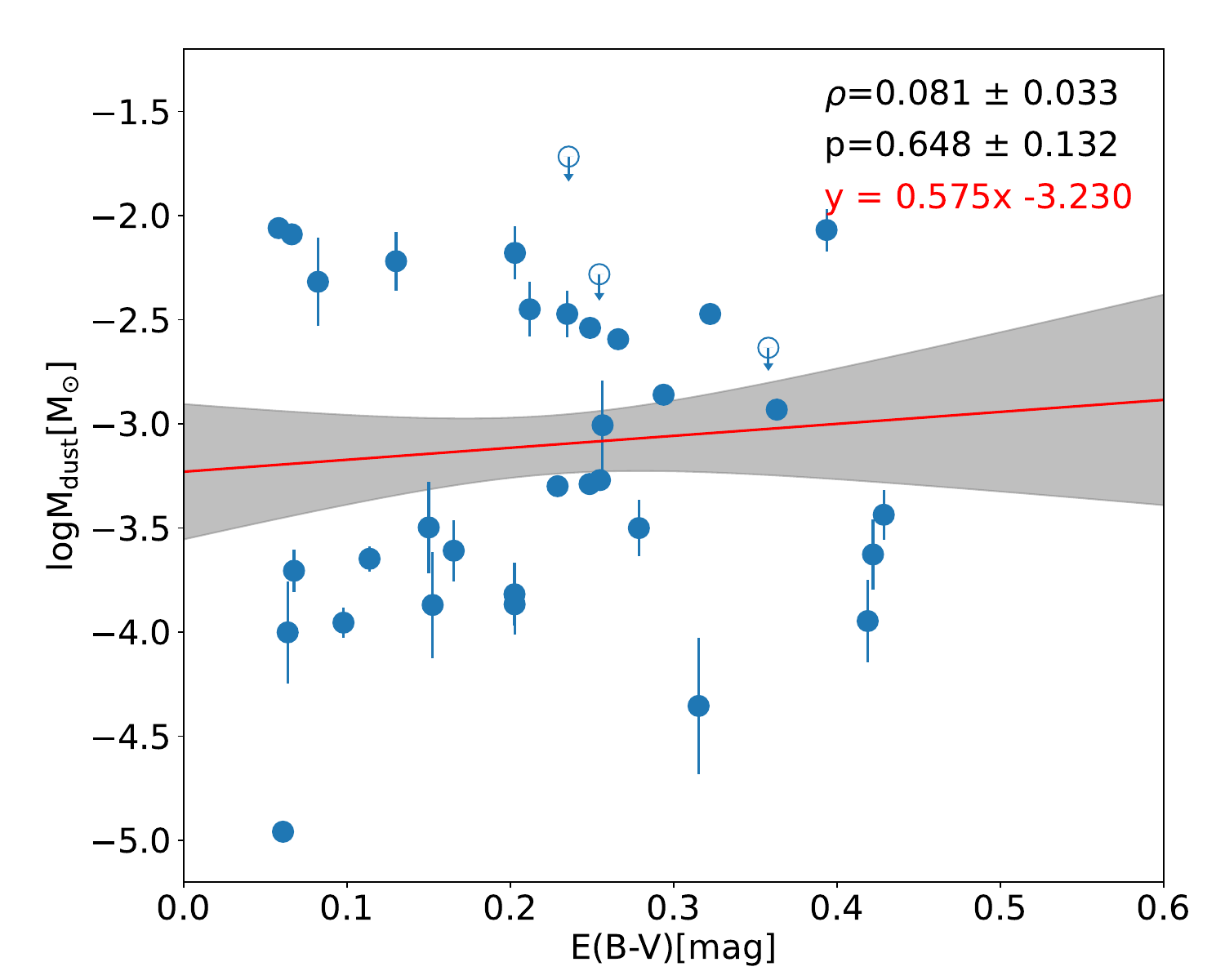}
\includegraphics[width=0.8\columnwidth]{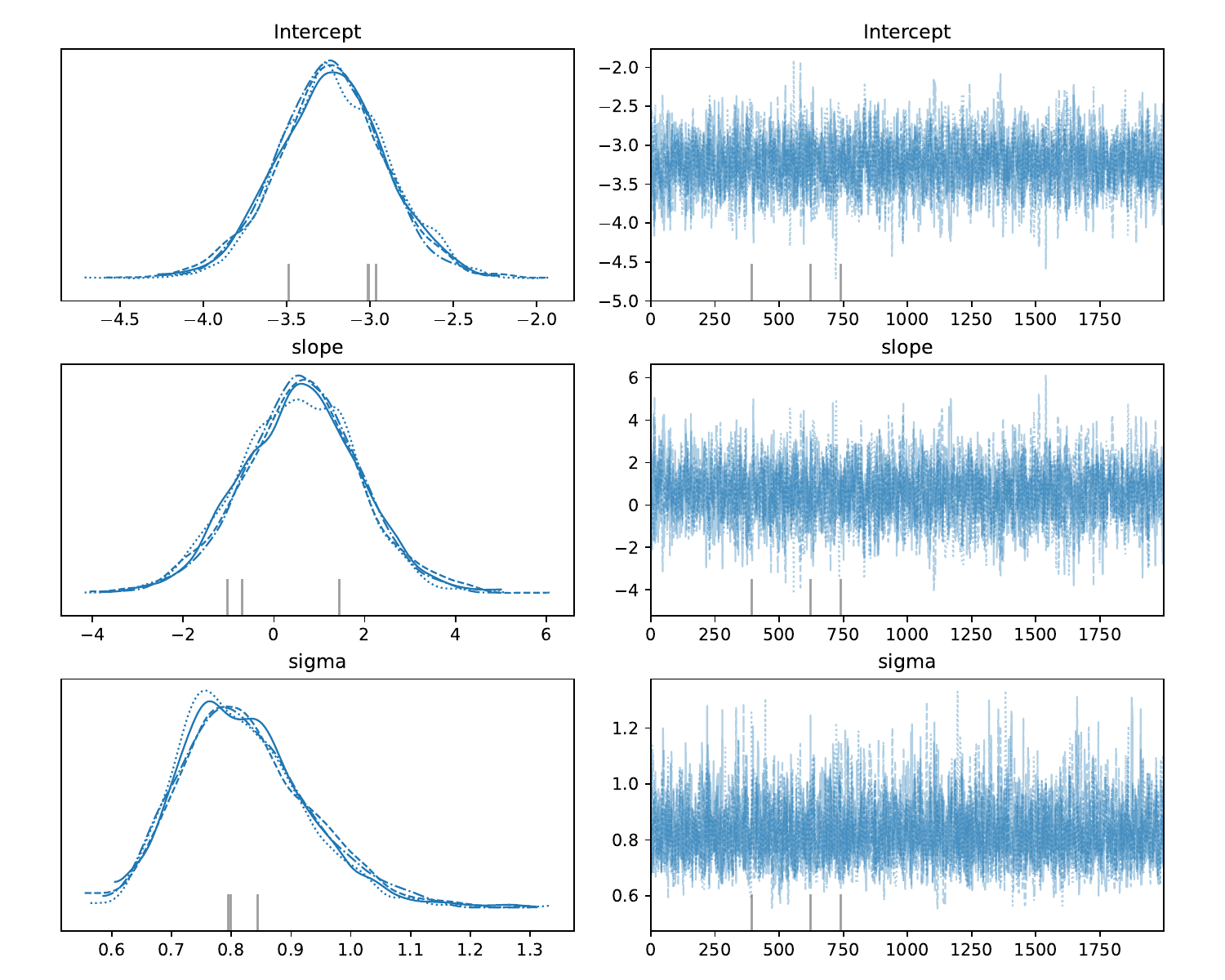}\\
\caption{The correlation between the largest dust mass M$_{\rm dust}$ of each SN labeled in bold in Table \ref{tab:sndata-photo} and the four environmental properties with EW(H$\alpha$), star formation rate intensity, oxygen abundance, and E(B-V). The left panels show the result of Bayesian regression in the red line with a 1-$\sigma$ uncertainty in the gray shades. Each Pearson correlation coefficient $\rho$ and $p$-value is shown along with its standard deviation. The posterior distributions of the fitting parameters are presented in the left panels.
\label{fig:correlation_Mdust}}
\end{figure*}

For each SN, we use the data at the epoch with the largest measured dust mass as their dust contribution labeled in bold in Table \ref{tab:sndata-photo}, and investigated whether there is any correlation between SN dust and host environmental properties. We employed \texttt{PyMC3} \citep{Salvatier2016} for a Bayesian regression assuming Gaussian uncertainties for the environmental parameters, and evaluated the Pearson correlation coefficient $\rho$ and $p$-value to determine the existence and the strength of correlations. Given the small number of the dusty SNe in H-poor and type IIn, we did not distinguish SN types and used the whole dusty SN sample to do the simulation.

Figure \ref{fig:correlation_Mdust} presents the result of the linear regression between the SN dust mass and the four environmental properties and posterior distribution of the fitting parameters. We note that one marginal correlation is a negative correlation between the dust mass and star formation rate, according to the Pearson test of correlation coefficient $\rho$ = -0.468 $\pm$ 0.057 and $p$-value = 0.049 $\pm$ 0.031. This means that SNe would be more mid-IR luminous and more dust-rich at the region of lower star formation rate. Unlike SFR, there is no correlation between EW(H$\alpha$) and SN dust mass, with a very low correlation coefficient $\rho$ = 0.002 $\pm$ 0.034 (p-value = 0.881 $\pm$ 0.085). In addition, it is worth noting that we did not find significant correlations between SN dust mass and  metallicity and host extinction with lower correlation coefficient $\rho$ = 0.058 $\pm$ 0.031 and 0.081 $\pm$ 0.033 (p-value = 0.725 $\pm$ 0.133 and 0.648 $\pm$ 0.132) respectively, which were thought to be key factors affecting the mass-loss history of progenitors and the CSM environment of SNe. It seems that the formation process of SN dust is less dependent on metallicities, which is consistent with the findings of \cite{Moriya2023} for type IIn SNe. The mid-IR emission used for the measurement of SN dust mass is insensitive to the interstellar extinction, and can be a reasonable explanation for this result. \\\\\\


\section{Conclusions} \label{sec:conclusions}

In this work, we discovered 42 mid-IR luminous dusty SNe with local IFS data based on {\it Spitzer} and {\it WISE} images. The observed mid-IR emission indicates the presence of newly formed dust, or pre-existing dust heated by the radiation from the SN or CSM interactions. We carried out a systematic analysis of the SN host environments and their dust properties, for understanding the dust-veiled exploding stars, and whether SN dust formation is associated with the local environments. Our main conclusions have been summarized as follows. 

\begin{enumerate}
  \item We note a clear difference in the host environments of dusty SNe compared to typical CCSN types. Dusty SNe prefer the locations with higher EW(H$\alpha$), lower metallicity, and heavier host extinctions, which indicate that they may come from younger and higher-mass progenitors. 

  \item Dusty SNe in different subtypes show diversities in terms of their host environmental properties. We found the same increasing sequence in the values of EW(H$\alpha$) and oxygen abundance from H-rich, type IIn to H-poor dusty SNe. These differences in subtypes of dusty SNe reflect the diversity of their progenitors. The progenitors of H-poor dusty SNe seem to be mostly sensitive with their host environments. 
  
  \item Based on the mid-IR fitting results, we found that the pre-exiting dust can dominate the mid-IR emission of CCSNe from quite early time around 100 days up to 1000 days post discovery, which become the main source of the observed dust in our sample CCSNe. 
  
  \item We found that one marginal correlation is a negative correlation between the dust mass and star formation rate. This means that SNe would be more mid-IR luminous and more dust-rich at the region with lower star formation rate. 
  
  \item We did not find significant correlations of the SN dust mass with the metallicity and the host extinction, which were thought to be key factors affecting the mass-loss history of progenitors and the CSM environment of SNe. This may indicate that the dust formation process in SNe might be insensitive to metallicity and the dust condition of the host regions. 


\end{enumerate}

Overall, our studies illustrate that the environmental studies of dusty SNe offer a powerful approach to investigate the dust formation in SNe and their potential correlation with progenitor systems. It is possible that some bias exists in our sample and affects our result. The SN types and dust measured in our sample are limited by {\it WISE} in observation band, resolution and epochs. Future wide-field IR surveys increasing the amount and diversity of dusty SNe with less biases, could allow for an even better statistical sample for a systematic study. We expect more observational data and theoretical models to improve our understanding of dusty SNe and their environments. The photometry and spectrometry data from {\it JWST} can offer unprecedented opportunity to resolve the details of the immediate environment of the dusty SNe, trace the evolution of dust from SN to SN remnant phase, and test more precise dust models with different components \cite{Sarangi2022,Shahbandeh2023,Zsiros2023}.

\begin{acknowledgments}
We thank the referee for valuable comments. We thank all the people that have made this paper. LX is thankful for the support from National Natural Science Foundation of China (grant No. 12103050 ), Advanced Talents Incubation Program of the Hebei University, and Midwest Universities Comprehensive Strength Promotion project. L.G. acknowledges financial support from AGAUR, CSIC, MCIN and AEI 10.13039/501100011033 under projects PID2023-151307NB-I00, PIE 20215AT016, CEX2020-001058-M, ILINK23001, COOPB2304, and 2021-SGR-01270.
Y.Y. appreciates the generous financial support provided to the supernova group at U.C. Berkeley by Gary and Cynthia Bengier, Clark and Sharon Winslow, Sanford Robertson, and numerous other donors. X.W. and Z.H. gratefully acknowledge the support provided by the NSFC under Grant No. 12288102.

This research is based on observations made with the Wide-field Infrared Survey Explorer ({\it WISE}) and the {\it Spitzer} Space Telescope, which are projects of the Jet Propulsion Laboratory, California Institute of Technology, and funded by the National Aeronautics and Space Administration and the National Science Foundation. We acknowledge the availability of unWISE services, which is a reprocessing of WISE data to improved depth and spatial resolution. This publication makes use of data products from the Centro Astronómico Hispano Alemán (CAHA) at Calar Alto, operated jointly by the Max-Planck-Institut für Astronomie (MPIA) and the Instituto de Astrofísica de Andalucía (CSIC), the European Southern Observatory under ESO programme(s): 096.D-0296 (A), 0103.D-0440 (A), 096.D-0263 (A), 097.B-0165 (A), 097.D-0408 (A), 0104.D-0503 (A), 60.A9301 (A), 096.D-0786 (A), 097.D-1054 (B), 0101.C-0329 (D), 0100.D-0341 (A), 1100.B-0651 (A), 094.B-0298 (A), 097.B-0640 (A), 0101.D-0748 (A), 095.D-0172 (A), 1100.B-0651 (A), 0100.D-0649 (F), 096.B-0309 (A), and SDSS-IV/MaNGA provided by the Alfred P. Sloan Foundation, the US Department of Energy Office of Science, and the Participating Institutions. SDSS-IV acknowledges support and resources from the Center for High Performance Computing at the University of Utah. 

\end{acknowledgments}

%

\vspace{5mm}
\facilities{{\it Spitzer/IRAC} \citep{Fazio2004}, {\it WISE} \citep{Wright2010}, {\it CAO:3.5m(PMAS/PPak)} \citep{Verheijen2004}, {\it VLT/MUSE} \citep{Bacon2014}, {\it APO:2.5m(SDSS)}} \citep[MaNGA;][]{Bundy2015}


\software{astropy \citep{2013A&A...558A..33A,2018AJ....156..123A},  
          numpy \citep{2013RMxAA..49..137F}, 
          SciPy \citep{2020SciPy-NMeth},
          Matplotlib \citep{1996A&AS..117..393B},
          Seaborns \citep{Waskom2021},
          Photutils \citep{larry_bradley_2023_7946442},
          SFFT \citep{sfft_zenodo}
          PyMC3 \citep{Salvatier2016}}



\appendix

\section{The redshift distribution of the IFS sample}
\begin{figure}[ht!]
\centering
\includegraphics[width=0.6\columnwidth]{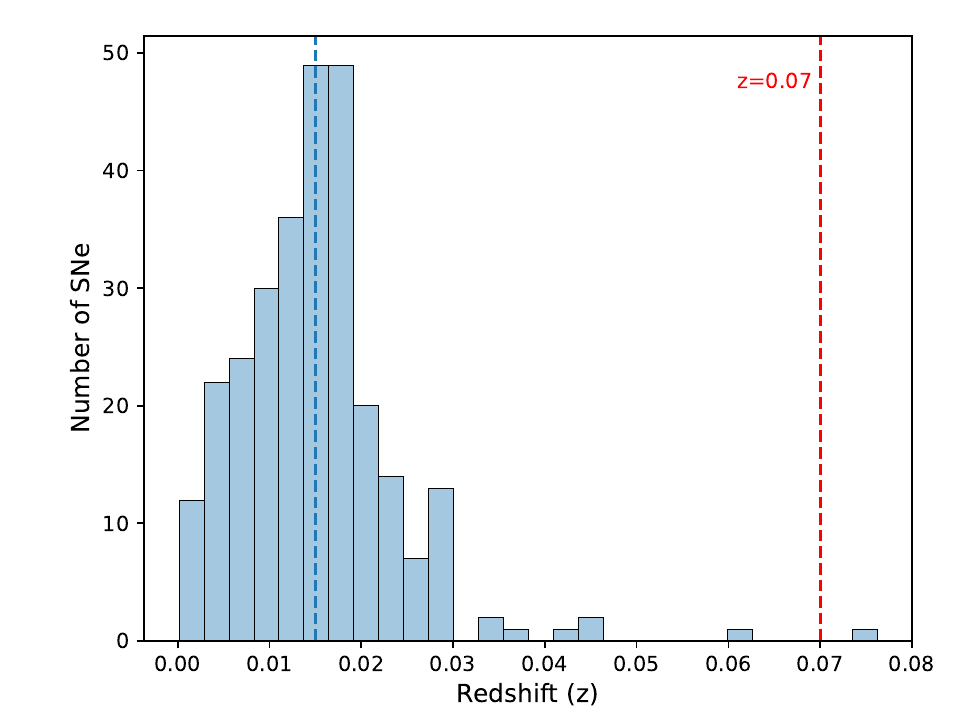}
\caption{The redshift distribution of the IFS CCSN sample as the discussion in Section \ref{sec:CCSNe-IFS}. The blue vertical line at z $=$ 0.015 is the median redshift of the CCSN IFS sample. The assumed detection limit of {\it WISE} is marked by the red vertical line of z $=$ 0.07 based on the observations of \cite{Myers2024}.
\label{fig:reshift-IFS}}
\end{figure}

\section{The mid-IR photometry and dust properties}\label{sec:photometry-WISE}
Here, we present the mid-IR photometry and SED fitting results of the dusty SNe in our sample in Table \ref{tab:sndata-photo}. The data at epoch labeled in bold is used as the dust contribution of the SNe. The photometry of SN 2003gd and SN 2013ej are from {\it Spitzer} at 3.5$\mu$m and 4.5$\mu$m. The other SNe are detected by {\it WISE} at 3.4$\mu$m and 4.6$\mu$m. There are four SNe (SN 2014ay, SN 2017gmr, SN 2018bbl, SN 2018cuf), which at some epochs are only positively detected at 4.6 $\mu$m, so we calculated the flux with a 3$\sigma$ upper limit at 3.4 $\mu$m and upper limits for their corresponding dust parameters.

\begin{deluxetable}{ccccccccc}
\tablecaption{The mid-IR photometry and SED fitting results. \label{tab:sndata-photo}}
\tablenum{3}
\tablecolumns{3}
\tablehead{
\colhead{Name} & \colhead{MJD} & \colhead{Epoch} & \colhead{f$_{\rm 3.4\mu m}$} & \colhead{f$_{\rm 4.6\mu m}$} & \colhead{R$_{\rm BB}$} &\colhead{M$_{\rm dust}$} & \colhead{T$_{\rm dust}$} & \colhead{L$_{\rm dust} $} \\
\colhead{ } & \colhead{ } &  \colhead{days} & \colhead{$\mu$Jy} & \colhead{$\mu$Jy} & \colhead{$10^{16}$cm} & \colhead{$10^{-5}$\msolar} & \colhead{K} & \colhead{$10^{6}$\lsolar }
}
\startdata
 \tableline
SN  2003gd$^{*}$	&	53211	&	\textbf{409}	&	20	$\pm$ 3 &	71	$\pm$ 14&	0.4			&	\textbf{1.1}			&	520 			&	0.1				\\
SN2010dr	&	55570	&	\textbf{220}	&	319	$\pm$	14	&	395	$\pm$	15	&	0.69	$\pm$	0.08	&	\textbf{53.7	$\pm$	11.2}	&	851 	$\pm$	41 	&	19.42		$\pm$	0.02		\\
SN 2013ej$^{*}$	&	56673	&	238	&	823	$\pm$ 99	&	1525	$\pm$	98 	&	1.7			&	62			&	490 			&	2.9			\\
	&	56862	&	\textbf{261}	&	630	$\pm$	75	&	1201	$\pm$ 55 &	1.6			&	\textbf{74}			&	465 			&	2.2	\\
	&	57036	&	439	&	41	$\pm$ 7	&	133	$\pm$	15	&	1.2			&	64			&	360 			&	0.4	\\
SN 2014az	&	56831	&	\textbf{54}	&	203	$\pm$	8	&	227	$\pm$	14	&	0.44	$\pm$	0.07	&	\textbf{22.5	$\pm$	6.4}	&	928 	$\pm$	60 	&	10.81		$\pm$	0.01 	\\
	&	57011	&	233	&	40	$\pm$	10	&	63	$\pm$	14	&	0.39	$\pm$	0.20	&	16.7	$\pm$	9.9	&	719 	$\pm$	162 	&	3.55		$\pm$	0.01	\\
SN 2014ay	&	56918	&	137	&	110	$\pm$	7	&	188	$\pm$	11	&	0.77	$\pm$	0.12	&	65.6	$\pm$	19.5	&	679 	$\pm$	37 	&	11.73		$\pm$	0.01 	\\
 	&	57101	&	\textbf{319}	&	$<$ 21	&	69	$\pm$	9	&	$<$ 1.47	&	$<$ \textbf{232}	&	$<$ 456  	&	$<$ 11.61\\
ASASSN-14az	&	56827	&	30	&	612	$\pm$	7	&	790	$\pm$	13	&	1.04	$\pm$	0.04	&	121.6	$\pm$	9.1	&	824 	$\pm$	13 	&	39.72		$\pm$	0.03 	\\
	&	57006	&	\textbf{208}	&	45	$\pm$	8	&	161	$\pm$	16	&	2.12	$\pm$	0.67	&	\textbf{481.1	$\pm$	301.6}	&	464 	$\pm$	39 	&	25.64		$\pm$	0.08 	\\
SN 2014cw	&	57166	&	\textbf{267}	&	38	$\pm$	10	&	70	$\pm$	16	&	0.54	$\pm$	0.28	&	\textbf{31.6	$\pm$	19.7}	&	644 	$\pm$	129 	&	4.81		$\pm$	0.01	\\
ASASSN-14ma	&	57198	&	\textbf{194}	&	41	$\pm$	9	&	60	$\pm$	15	&	0.35	$\pm$	0.21	&	\textbf{13.5	$\pm$	15.9}	&	749 	$\pm$	176 	&	3.27		$\pm$	0.01	\\
ASASSN-15ab	&	57230	&	\textbf{205}	&	51	$\pm$	10	&	188	$\pm$	18	&	2.37	$\pm$	0.78	&	\textbf{604.0	$\pm$	394.8}	&	459 	$\pm$	41 	&	31.27		$\pm$	0.10 	\\
	&	57419	&	394	&	34	$\pm$	9	&	126	$\pm$	17	&	1.99	$\pm$	0.92	&	424.4	$\pm$	391.2	&	456 	$\pm$	57 	&	21.31		$\pm$	0.03 	\\
ASASSN-15jp	&	57369	&	\textbf{206}	&	59	$\pm$	9	&	91	$\pm$	15	&	0.46	$\pm$	0.19	&	\textbf{23.6	$\pm$	18.5}	&	726 	$\pm$	109 	&	5.18		$\pm$	0.01 	\\
ASASSN-15kj	&	57378	&	204	&	1220	$\pm$	12	&	1450	$\pm$	15	&	1.22	$\pm$	0.03	&	170.5	$\pm$	8.5	&	884 	$\pm$	10 	&	69.92		$\pm$	0.05 	\\
	&	57568	&	\textbf{395}	&	995	$\pm$	16	&	1370	$\pm$	13	&	1.51	$\pm$	0.05	&	\textbf{254.7	$\pm$	15.3}	&	785 	$\pm$	11 	&	71.33		$\pm$	0.06 \\
	&	57738	&	565	&	696	$\pm$	12	&	1020	$\pm$	16	&	1.45	$\pm$	0.06	&	232.3	$\pm$	18.8	&	748 	$\pm$	12 	&	56.34		$\pm$	0.05 \\
	&	57934	&	760	&	305	$\pm$	12	&	528	$\pm$	15	&	1.33	$\pm$	0.11	&	193.2	$\pm$	30.8	&	673 	$\pm$	20 	&	33.53		$\pm$	0.02 \\
	&	58099	&	925	&	180	$\pm$	10	&	327	$\pm$	16	&	1.13	$\pm$	0.15	&	138.7	$\pm$	35.1	&	652 	$\pm$	30 	&	21.92		$\pm$	0.02 \\
ASASSN-15ln	&	57389	&	\textbf{196}	&	69	$\pm$	7	&	91	$\pm$	14	&	0.37	$\pm$	0.13	&	\textbf{15.2	$\pm$	10.5}	&	807 	$\pm$	114 	&	4.64		$\pm$	0.01	\\
ASASSN-15ng	&	57421	&	\textbf{194}	&	58	$\pm$	9	&	178	$\pm$	13	&	1.77	$\pm$	0.45	&	\textbf{337.2	$\pm$	171.5}	&	498 	$\pm$	37 	&	22.63		$\pm$	0.03 	\\
SN 2015bj	&	57576	&	\textbf{227}	&	57	$\pm$	9	&	77	$\pm$	10	&	0.35	$\pm$	0.12	&	\textbf{13.6	$\pm$	9.2}	&	795 	$\pm$	117 	&	3.96		$\pm$	0.01	\\
SN 2016B	&	57543	&	152	&	459	$\pm$	9	&	687	$\pm$	15	&	1.22	$\pm$	0.07	&	164.4	$\pm$	17.4	&	739 	$\pm$	15 	&	38.23		$\pm$	0.04 	\\
	&	57742	&	\textbf{351}	&	36	$\pm$	10	&	137	$\pm$	21	&	2.12	$\pm$	1.04	&	\textbf{481.4	$\pm$	472.0}	&	453 	$\pm$	59 	&	23.74		$\pm$	0.03 	\\
SN 2016C	&	57578	&	\textbf{187}	&	63	$\pm$	10	&	119	$\pm$	19	&	0.72	$\pm$	0.29	&	\textbf{55.8	$\pm$	44.3}	&	639 	$\pm$	87 	&	8.24		$\pm$	0.01	\\
SN 2016X	&	57560	&	152	&	513	$\pm$	8	&	487	$\pm$	14	&	0.48	$\pm$	0.04	&	28.5	$\pm$	3.9	&	1092 	$\pm$	40 	&	24.18		$\pm$	0.01 	\\
	&	57756	&	349	&	142	$\pm$	10	&	230	$\pm$	17	&	0.79	$\pm$	0.15	&	69.3	$\pm$	25.4	&	702 	$\pm$	48 	&	13.72		$\pm$	0.01 	\\
	&	57920	&	\textbf{513}	&	43	$\pm$	7	&	169	$\pm$	15	&	2.48	$\pm$	0.73	&	\textbf{661.7	$\pm$	390.4}	&	446 	$\pm$	35 	&	30.81		$\pm$	0.04 \\
SN 2016aqf	&	57646	&	\textbf{202}	&	101	$\pm$	4	&	113	$\pm$	8	&	0.31	$\pm$	0.05	&	\textbf{11.1	$\pm$	3.7}	&	929 	$\pm$	71 	&	5.38		$\pm$	0.01	\\
SN 2016bkv	&	57708	&	239	&	221	$\pm$	13	&	218	$\pm$	19	&	0.35	$\pm$	0.08	&	14.5	$\pm$	6.0	&	1047 	$\pm$	113 	&	10.52		$\pm$	0.01 	\\
	&	57866	&	397	&	321	$\pm$	13	&	329	$\pm$	24	&	0.46	$\pm$	0.08	&	25.0	$\pm$	8.4	&	1005 	$\pm$	83 	&	15.71		$\pm$	0.01 	\\
	&	58070	&	\textbf{601}	&	133	$\pm$	16	&	186	$\pm$	20	&	0.57	$\pm$	0.16	&	\textbf{36.6	$\pm$	20.2}	&	775 	$\pm$	89 	&	9.84		$\pm$	0.01 \\
SN 2016blz	&	57599	&	\textbf{111}	&	162	$\pm$	7	&	303	$\pm$	12	&	1.12	$\pm$	0.11	&	\textbf{138	$\pm$	26.5}	&	643 	$\pm$	22 	&	20.82		$\pm$	0.01 	\\
	&	57798	&	309	&	102	$\pm$	7	&	149	$\pm$	12	&	0.54	$\pm$	0.10	&	32.8	$\pm$	9.9	&	754 	$\pm$	54 	&	8.09		$\pm$	0.008 	\\
SN 2016bmi	&	57647	&	\textbf{154}	&	217	$\pm$	9	&	384	$\pm$	16	&	1.17	$\pm$	0.12	&	\textbf{150.0	$\pm$	30.1}	&	663 	$\pm$	24 	&	24.92		$\pm$	0.03 	\\
	&	57839	&	347	&	67	$\pm$	8	&	120	$\pm$	13	&	0.67	$\pm$	0.19	&	49.6	$\pm$	27	&	655 	$\pm$	63 	&	7.94		$\pm$	0.01 	\\
SN 2016cdd	&	57580	&	\textbf{58}	&	267	$\pm$	7	&	342	$\pm$	13	&	0.68	$\pm$	0.06	&	\textbf{51.4	$\pm$	9.0}	&	829 	$\pm$	31 	&	17.13		$\pm$	0.01 	\\
	&	57775	&	253	&	76	$\pm$	8	&	120	$\pm$	13	&	0.58	$\pm$	0.15	&	34.2	$\pm$	17.6	&	710 	$\pm$	69 	&	7.02		$\pm$	0.01 	\\
	&	57942	&	420	&	23	$\pm$	7	&	55	$\pm$	13	&	0.71	$\pm$	0.46	&	54	$\pm$	47.6	&	556 	$\pm$	113 	&	5.17		$\pm$	0.01 	\\
SN 2016cvk	&	57887	&	\textbf{335}	&	31	$\pm$	10	&	61	$\pm$	16	&	0.54	$\pm$	0.33	&	\textbf{31.8	$\pm$	32.2}	&	626 	$\pm$	144 	&	4.41		$\pm$	0.01	\\
SN 2016cyx	&	57768	&	\textbf{192}	&	81	$\pm$	12	&	116	$\pm$	15	&	0.47	$\pm$	0.16	&	\textbf{24.6	$\pm$	16.7}	&	762 	$\pm$	106 	&	6.26		$\pm$	0.01 	\\
SN 2016gkg	&	57740	&	89	&	173	$\pm$	10	&	302	$\pm$	15	&	1.02	$\pm$	0.14	&	114.7	$\pm$	29.7	&	667 	$\pm$	32 	&	19.4		$\pm$	0.01 	\\
	&	57935	&	284	&	178	$\pm$	8	&	264	$\pm$	13	&	0.75	$\pm$	0.09	&	61.5	$\pm$	14.9	&	744 	$\pm$	35 	&	14.61		$\pm$	0.01 	\\
	&	58101	&	449	&	145	$\pm$	10	&	267	$\pm$	16	&	1.03	$\pm$	0.16	&	116.3	$\pm$	35.7	&	648 	$\pm$	35 	&	18.12		$\pm$	0.01 	\\
	&	58302	&	650	&	98	$\pm$	8	&	242	$\pm$	18	&	1.52	$\pm$	0.28	&	251.1	$\pm$	90.2	&	550 	$\pm$	30 	&	23.35		$\pm$	0.07 	\\
	&	58464	&	\textbf{813}	&	38	$\pm$	9	&	124	$\pm$	19	&	1.64	$\pm$	0.75	&	\textbf{290.0	$\pm$	262.4}	&	482 	$\pm$	60 	&	17.54		$\pm$	0.02 \\
	&	58667	&	1016	&	44	$\pm$	8	&	114	$\pm$	16	&	1.1	$\pm$	0.42	&	131.4	$\pm$	100.3	&	540 	$\pm$	63 	&	11.51		$\pm$	0.01 	\\
	&	58828	&	1177	&	21	$\pm$	7	&	38	$\pm$	12	&	0.37	$\pm$	0.29	&	15.2	$\pm$	23.6	&	658 	$\pm$	180 	&	2.48		$\pm$	0.01 \\
\tableline
\enddata
\end{deluxetable}

\begin{deluxetable}{ccccccccc}
\tablecaption{Continued \label{tab:sndata-photo}}
\tablenum{3}
\tablecolumns{3}
\tablehead{
\colhead{Name} & \colhead{MJD} & \colhead{Epoch} & \colhead{f$_{\rm 3.4\mu m}$} & \colhead{f$_{\rm 4.6\mu m}$} & \colhead{R$_{\rm BB}$} &\colhead{M$_{\rm dust}$} & \colhead{T$_{\rm dust}$} & \colhead{L$_{\rm dust} $} \\
\colhead{ } & \colhead{ } &  \colhead{days} & \colhead{$\mu$Jy} & \colhead{$\mu$Jy} & \colhead{$10^{16}$cm} & \colhead{$10^{-5}$\msolar} & \colhead{K} & \colhead{$10^{6}$\lsolar }
}
\startdata
 \tableline
SN 2016hgm	&	57947	&	\textbf{267}	&	47	$\pm$	11	&	111	$\pm$	20	&	0.96	$\pm$	0.47	&	\textbf{98.5	$\pm$	96.5}	&	566 	$\pm$	88 	&	9.98		$\pm$	0.01 	\\
SN 2016hwn	&	57902	&	\textbf{203}	&	54	$\pm$	8	&	102	$\pm$	14	&	0.66	$\pm$	0.23	&	\textbf{47.5	$\pm$	33.1}	&	639 	$\pm$	77 	&	7.04		$\pm$	0.01		\\
SN 2016iae	&	57783	&	\textbf{84}	&	285	$\pm$	21	&	578	$\pm$	27	&	1.76	$\pm$	0.25	&	\textbf{336.2	$\pm$	93.9}	&	612 	$\pm$	30 	&	43.5		$\pm$	0.02 	\\
SN 2017bzb	&	58061	&	\textbf{242}	&	247	$\pm$	7	&	653	$\pm$	14	&	2.75	$\pm$	0.15	&	\textbf{813.3	$\pm$	90.4}	&	533 	$\pm$	9 	&	68.2		$\pm$	0.09 	\\
SN 2017faa	&	58128	&	197	&	294	$\pm$	10	&	359	$\pm$	17	&	0.64	$\pm$	0.07	&	46.6	$\pm$	9.93	&	861 	$\pm$	40 	&	17.53		$\pm$	0.01 	\\
	&	58287	&	\textbf{355}	&	355	$\pm$	12	&	422	$\pm$	13	&	0.66	$\pm$	0.06	&	\textbf{50.2	$\pm$	8.24}	&	882 	$\pm$	34 	&	20.44		$\pm$	0.02 \\
	&	58493	&	562	&	123	$\pm$	13	&	177	$\pm$	18	&	0.59	$\pm$	0.16	&	38.5	$\pm$	19.6	&	757 	$\pm$	79 	&	9.64		$\pm$	0.01 	\\
	&	58654	&	723	&	60	$\pm$	9	&	91	$\pm$	14	&	0.46	$\pm$	0.18	&	22.9	$\pm$	17.7	&	731 	$\pm$	110 	&	5.14		$\pm$	0.01 \\
SN 2017fbu	&	58124	&	\textbf{187}	&	47	$\pm$	9	&	74	$\pm$	15	&	0.42	$\pm$	0.21	&	\textbf{19.7	$\pm$	9.3}	&	720 	$\pm$	138 	&	4.23		$\pm$	0.01	\\
SN 2017ffq	&	58195	&	251	&	102	$\pm$	11	&	88	$\pm$	13	&	0.17	$\pm$	0.07	&	3.7	$\pm$	2.8	&	1223 	$\pm$	276 	&	4.76		$\pm$	0.01	\\
	&	58359	&	\textbf{415}	&	6	$\pm$	2	&	11	$\pm$	3	&	0.20	$\pm$	0.15	&	\textbf{4.4	$\pm$	6.7}	&	662 	$\pm$	179 	&	0.73		$\pm$	0.01	\\
SN 2017fqk	&	58132	&	\textbf{173}	&	39	$\pm$	11	&	55	$\pm$	19	&	0.32	$\pm$	0.24	&	\textbf{11.3	$\pm$	10.3}	&	768 	$\pm$	242 	&	2.95		$\pm$	0.01	\\
SN 2017gax	&	58124	&	\textbf{145}	&	143	$\pm$	10	&	264	$\pm$	10	&	1.03	$\pm$	0.13	&	\textbf{117	$\pm$	28.8}	&	646 	$\pm$	30 	&	17.91		$\pm$	0.01 	\\
	&	58327	&	347	&	34	$\pm$	11	&	56	$\pm$	11	&	0.40	$\pm$	0.25	&	17.9	$\pm$	21.7	&	692 	$\pm$	166 	&	3.39		$\pm$	0.01	\\
SN 2017gmr	&	58124	&	\textbf{124}	&	1300	$\pm$	13	&	1790	$\pm$	15	&	1.74	$\pm$	0.04	&	\textbf{336.8	$\pm$	15.1}	&	783 	$\pm$	8 	&	93.54		$\pm$	0.08 	\\
	&	58324	&	324	&	134	$\pm$	11	&	244	$\pm$	15	&	0.98	$\pm$	0.17	&	103.9	$\pm$	35.4	&	652 	$\pm$	41 	&	16.37		$\pm$	0.02 	\\
	&	58485	&	484	&	$<$ 39	&	89	$\pm$	16	&	$<$ 0.84	&	$<$ 77.1	&	$<$ 568  	&	$<$ 7.89	\\
SN 2017hcc	&	58287	&	259	&	764	$\pm$	10	&	1060	$\pm$	21	&	1.36	$\pm$	0.06	&	207.0	$\pm$	18.3	&	777 	$\pm$	14 	&	56.13		$\pm$	0.05 	\\
	&	58450	&	421	&	1550	$\pm$	11	&	1480	$\pm$	17	&	0.85	$\pm$	0.03	&	87.2	$\pm$	4.7	&	1088 	$\pm$	16 	&	72.62		$\pm$	0.04 	\\
	&	58651	&	623	&	1070	$\pm$	12	&	1290	$\pm$	14	&	1.18	$\pm$	0.04	&	159.6	$\pm$	9.2	&	872 	$\pm$	12 	&	62.63		$\pm$	0.05 	\\
	&	58814	&	785	&	800	$\pm$	8	&	1080	$\pm$	15	&	1.3	$\pm$	0.04	&	190.6	$\pm$	11.7	&	796 	$\pm$	10 	&	55.72		$\pm$	0.05 	\\
	&	59019	&	\textbf{990}	&	413	$\pm$	9	&	733	$\pm$	18	&	1.63	$\pm$	0.10	&	\textbf{289.3	$\pm$	33.3}	&	662 	$\pm$	13 	&	47.74		$\pm$	0.05 	\\
	&	59179	&	1151	&	323	$\pm$	7	&	605	$\pm$	13	&	1.6	$\pm$	0.08	&	279.8	$\pm$	29.2	&	641 	$\pm$	11 	&	41.72		$\pm$	0.02 	\\
	&	59383	&	1354	&	180	$\pm$	8	&	355	$\pm$	20	&	1.32	$\pm$	0.175	&	189.4	$\pm$	49.7	&	622 	$\pm$	27 	&	25.83		$\pm$	0.01 	\\
SN 2017hpi	&	58221	&	169	&	221	$\pm$	9	&	433	$\pm$	15	&	1.45	$\pm$	0.13	&	227.8	$\pm$	40.4	&	624 	$\pm$	19 	&	31.41		$\pm$	0.02 	\\
	&	58431	&	\textbf{378}	&	43	$\pm$	7	&	144	$\pm$	13	&	1.82	$\pm$	0.55	&	\textbf{355.2	$\pm$	213.4}	&	477 	$\pm$	40 	&	20.84		$\pm$	0.02 	\\
SN 2017ivv	&	58240	&	\textbf{141}	&	141	$\pm$	7	&	439	$\pm$	16	&	2.85	$\pm$	0.28	&	\textbf{871.2	$\pm$	171.8}	&	494 	$\pm$	14 	&	57.12		$\pm$	0.06 	\\
	&	58404	&	305	&	31	$\pm$	6	&	61	$\pm$	12	&	0.51	$\pm$	0.25	&	32.8	$\pm$	23.3	&	622 	$\pm$	97 	&	4.46		$\pm$	0.01	\\
SN 2018pq	&	58304	&	\textbf{147}	&	63	$\pm$	9	&	239	$\pm$	12	&	2.82	$\pm$	0.67	&	\textbf{852.8	$\pm$	402.2}	&	452 	$\pm$	30 	&	41.65		$\pm$	0.05 	\\
SN 2018bbl	&	58423	&	\textbf{202}	&	$<$ 24	&	125	$\pm$	14	&	$<$ 4.23	&	$<$ \textbf{1919.5}	&	$<$ 373	&	$<$ 47.34 \\
SN 2018cuf	&	58594	&	\textbf{301}	&	$<$ 18	&	72	$\pm$	14	&	$<$ 2.21	&	$<$ \textbf{522.8} 	& $<$ 410	&	$<$ 18.11\\
SN 2020tlf	&	59382	&	273	&	29	$\pm$	7	&	42	$\pm$	12	&	0.29	$\pm$	0.20	&	9.0	$\pm$	12.2	&	758 	$\pm$	203 	&	2.26		$\pm$	0.01	\\
	&	59587	&	\textbf{479}	&	37	$\pm$	9	&	52	$\pm$	11	&	0.30	$\pm$	0.17	&	\textbf{10.0	$\pm$	11.2}	&	779 	$\pm$	182 	&	2.72		$\pm$	0.01	\\
\tableline
\enddata
\tablecomments{The data at the epoch in bold is used as the dust contribution of each SN in Figure \ref{fig:correlation_Mdust}. The dust type is classified into pre-exiting (pre) and newly-formed (nf) dust. *The photometry of SN 2003gd and SN 2013ej are from {\it Spitzer} at 3.5$\mu$m and 4.5$\mu$m, and the corresponding dust parameters are directly from \cite{Szalai2019}.} (This table is available in machine-readable form.)\\
\end{deluxetable}


\bibliography{sample631}{}
\bibliographystyle{aasjournal}



\end{document}